\documentclass[aps,showpacs,pre,onecolumn]{revtex4-1}

\usepackage[T1]{fontenc}
\usepackage[latin1]{inputenc}
\usepackage{amsmath}
\usepackage{amsthm}
\usepackage{amssymb}
\usepackage{subfigure}	
\usepackage{graphicx}
\usepackage{epstopdf}
\usepackage{color}
\usepackage[colorlinks=true,linkcolor=blue,citecolor=blue]{hyperref}
\usepackage{extramarks}
\usepackage{ae}
\usepackage{color}

\begin{document}
\date{}
\title{Periodic traveling waves in the $\phi^4$ model: Instability, stability and localized structures}
\author{Meng-Meng Liu}
\affiliation{School of Mathematics and Physics, University of Science\\
  and Technology Beijing, Beijing 100083, China}

\author{ Wen-Rong Sun}
\email{Corresponding author: sunwenrong@ustb.edu.cn}
\affiliation{School of Mathematics and Physics, University of Science\\
  and Technology Beijing, Beijing 100083, China\\
}

\author{Lei Liu}
\affiliation{College of Mathematics and Statistics, Chongqing University, Chongqing, 401331, China}

\author{P.G. Kevrekidis}
\affiliation{Department of Mathematics and Statistics, University of
  Massachusetts, Amherst, Massachusetts 01003-4515, USA}

\author{Lei Wang}
\affiliation{School of Mathematics and Physics, North China Electric Power University, Beijing, 102206, China}



\begin{abstract}
We consider the instability and stability of periodic stationary
solutions to the classical $\phi^4$ equation numerically. In
the superluminal regime,
the model possesses  dnoidal and cnoidal waves. The former are modulationally
unstable and the spectrum forms a figure eight intersecting at the
origin of the spectral plane. The latter can be modulationally stable and the spectrum
near the origin in that case is represented by  vertical bands along the purely
imaginary axis.
The instability of the cnoidal states in that case stems from
elliptical bands of complex eigenvalues far from the spectral plane origin.
In the subluminal regime, there exist only snoidal
waves which are modulationally unstable.
Considering the subharmonic perturbations, we show that the snoidal waves in the subluminal regime are spectrally unstable with respect to
all subharmonic perturbations, while for the dnoidal and cnoidal waves
in the superluminal regime, the transition between the spectrally
stable state and the spectrally unstable state occurs through a
Hamiltonian Hopf bifurcation. The dynamical evolution of the unstable
states is also considered, leading to some interesting  localization events on the spatio-temporally backgrounds.
 \\\\
\emph{Keywords: } Modulation instability; Spectral stability; Elliptic solutions; Rogue waves; Subharmonic perturbations. 
\end{abstract}

\maketitle

\newpage

\noindent\textbf{\large 1. Introduction}\\\hspace*{\parindent}

One of the most fundamental models in classical $1+1$-dimensional
field theory is the so-called $\phi^4$ model that has been summarized
in
recent reviews and books~\cite{b0,b11}.
The corresponding dynamical equation reads:
\begin{equation}\label{eqm}
\phi_{t t}-\phi_{x x}-\phi+\phi^{3}=0,
\end{equation}
and admits the following Hamiltonian:
\begin{equation}
H=\int \mathrm{d} x \left[\frac{1}{2}\left(\frac{\partial \phi}{\partial t}\right)^{2}+\frac{1}{2}\left(\frac{\partial \phi}{\partial x}\right)^{2}+\frac{1}{4}\left(\phi^{2}-1\right)^{2}\right].
\end{equation}
Here $\phi(x,t)$ is a real-valued function depending on the temporal
variable $t$ and the spatial variable $x$.  For  the $\phi^4$
model~(\ref{eqm})
(and related ones such as the sine-Gordon equation), there have been
extensive studies in the physics literature in the context of cosmological physics~\cite{b1}, condensed matter physics~\cite{b2},  statistical physics~\cite{b3,b4,b5,b6} and biophysics~\cite{b7}.
Furthermore, in quantum field theory, the $\phi^4$ model was used to study the transition between perturbative and non-perturbative sectors~\cite{b8,b9} and could be used to describe quantum mechanical instanton transitions in a double-well potential~\cite{b10}.
One of the pioneers of the field has recently
provided an  overview of the $\phi^4$ model by
 reviewing its applications to physics~\cite{b11b}. Importantly, the
 model
 has seen a resurgence of interest in its soliton and multi-soliton
 solutions over the past couple of years~\cite{manton1,manton2}.

 Historically, the comparison between the integrable
 sine-Gordon model and the non-integrable
 $\phi^4$ model~\cite{b0} has been one of the epicenters of the effort
 to appreciate the key differences between completely integrable and
 non-integrable classical field
 theories~\cite{b12,b13,b14,b15,b16,b17,b18}.  Therefore, some
 important topics discussed in the non-integrable $\phi^4$ theory are
 motivated by the same topics in the integrable sine-Gordon theory and
 relevant similarities and differences are assessed. 
For example, the remarkable complete integrability of the sine-Gordon
equation allows one to obtain analytic $N$-soliton solutions by using the
inverse scattering
transform~\cite{b19}.  $N$-soliton solutions allow one to
investigate the kink-kink interactions, kink-antikink interactions and
breathers of the  sine-Gordon equation analytically; see,
e.g.,~\cite{manton}
for kink interactions and also~\cite{avadh,hiroshima} for breather
interactions. Correspondingly, the  dynamics resulting from such
interactions have also been extensively studied in the $\phi^4$
theory~\cite{b12,b13,b14,b15,b16,b17,b18}.

However, there have been relatively few studies of stability of
periodic traveling
wave solutions to~(\ref{eqm}). Recently the work of~\cite{b21} has
proved the existence of three different forms of periodic traveling
wave solutions to~(\ref{eqm}). These solutions are given in
Propositions 3.1 (dnoidal waves), 3.2 (cnoidal waves) and 3.3 (snoidal
waves) of~\cite{b21}. The same work included a proof of the orbital
instability of
snoidal waves with respect to co-periodic perturbations~\cite{b21}.
It is important to note here that the latter waveform exists only for
subluminal speeds (i.e., for $c^2<1$), while the cnoidal and dnoidal
waveforms are only present for superluminal speeds (i.e., for $c^2>1$).
These results prompted the further consideration
of such solutions in the case of the $\phi^4$ model and of
their numerical stability. Indeed, our more concrete motivations are
as follows:

(1) For the sine-Gordon equation, spectral stability and modulational
instability
of periodic solutions have been studied~\cite{b22,b23,b24,b25}.
More recently, using the integrability of the relevant model, the work
of~\cite{b26} has obtained rogue-wave solutions in this setting which
describe localized structures on the background of librational
traveling waves. These results imply that rogue waves can be generated
from the modulationally unstable background, while if the periodic
traveling wave background is modulationally stable, the solutions are
not fully localized in space and time~\cite{b26,b27} and, thus, indeed
such localized rogue wave patterns are absent.
Since the existence of subluminal and superluminal solutions is
{\it shared} by the $\phi^4$ equation, it is natural to inquire
whether the $\phi^4$ equation also features modulational instability
of the periodic stationary solutions and whether such
localized structures (and possible rogue waves) could be generated on such modulational unstable backgrounds.
Since the $\phi^4$ equation is not integrable, the analytical methods
relying on the integrability can no longer be applied to generate
rogue-wave solutions here. Here we consider the modulational
instability of the $\phi^4$ equation numerically and explore the
its dynamical outcomes via direct numerical simulations.

(2) As stated before, the work of~\cite{b21} has proved the orbital
instability of snoidal waves to~(\ref{eqm}) with respect to co-periodic
perturbations. Yet,  the stability and instability of cnoidal and
dnoidal solutions of the $\phi^4$ model have not been investigated, to the
best
of our knowledge. Besides, we consider the subharmonic perturbations: perturbations that are periodic with period equal to an integer multiple of the period of the underlying solution.
In fact, using the integrability and considering the subharmonic
perturbations,
recent research efforts have considered the spectral and
orbital stability of certain integrable systems, such as the KdV
equation~\cite{b28,b29}, the NLS equation~\cite{b30,b31,b32},  the
modified KdV equation~\cite{b33}, the sine-Gordon equation~\cite{b34}
and the sinh-Gordon equation~\cite{b35}.  In 2017, the work
of~\cite{b34}
presented an analysis of the stability spectrum for all stationary
periodic solutions to the sine-Gordon equation using
integrability-based methods.
As discussed above the non-integrability of the $\phi^4$ model does
not
allow for an extension of such methods, hence we consider the
stability problem
in the latter case numerically using the so-called Hill's method~\cite{b36}.

A brief summary of the key numerical findings of this work are as follows.

(1) In the superluminal regime, we find that the dnoidal waves are
modulationally
unstable and their linearization spectrum forms a figure eight intersecting the
origin, while the cnoidal waves are modulationally stable and the
spectrum near the origin is represented by  vertical bands along
the purely imaginary axis. In the subluminal regime, the snoidal waves
are modulationally unstable since their spectrum forms a figure eight
intersecting at the origin. Since modulational instability is key
towards the formation of rogue waves, we numerically explore the relevant
dynamical instability and interestingly identify some
 spatio-temporally localized structures on the modulationally unstable backgrounds.
It is, once again, important to appreciate here that contrary to the
integrable sine-Gordon case (where the periodicity of the potential
allows for cnoidal/snoidal and dnoidal
solutions both for the
subliminal
and superluminal regime)~\cite{b24}, here cnoidal and dnoidal solutions are only
available in the superluminal regime and snoidal ones in the
subluminal case.

(2) Considering  subharmonic perturbations more specifically, we show that the snoidal waves in the subluminal regime are spectrally unstable with respect to
all subharmonic perturbations. For dnoidal and cnoidal waves in
the superluminal regime, we find that for such perturbations
a transition between the spectrally stable
state and the spectrally unstable state occurs. We explain this
through a Hamiltonian Hopf bifurcation.   For cnoidal and dnoidal
waves, the instability manifests itself when two imaginary eigenvalues
collide along the imaginary axis in a Hamiltonian Hopf bifurcation and
enter the right and left half planes along the figure eight (given the
Hamiltonian symmetry of the problem).  Stability, on the other hand,
arises through a Hamiltonian Hopf bifurcation in which two complex
conjugate pairs of eigenvalues come together onto the imaginary
axis. For snoidal waves, we show that there are two eigenvalues
fixed on the real axis, which implies that the snoidal waves in the
subluminal regime are spectrally unstable with respect to
all subharmonic perturbations.


The paper is organized as follows. The modulational
stability/instability and spatio-temporally localized structures are studied numerically
in Section 2. In Section 3, we study the stability and instability of
elliptic solutions with respect to subharmonic perturbations using
Hill's
method.  Section 4 summarizes our findings and offers some directions
for future study.
\\
\\
\noindent\textbf{\large 2. Localized Structures and their Modulational
Stability} \\\hspace*{\parindent}

\subsection{Review of Existence Results}

\begin{figure}
  \hspace{-1cm}\includegraphics[scale=0.400]{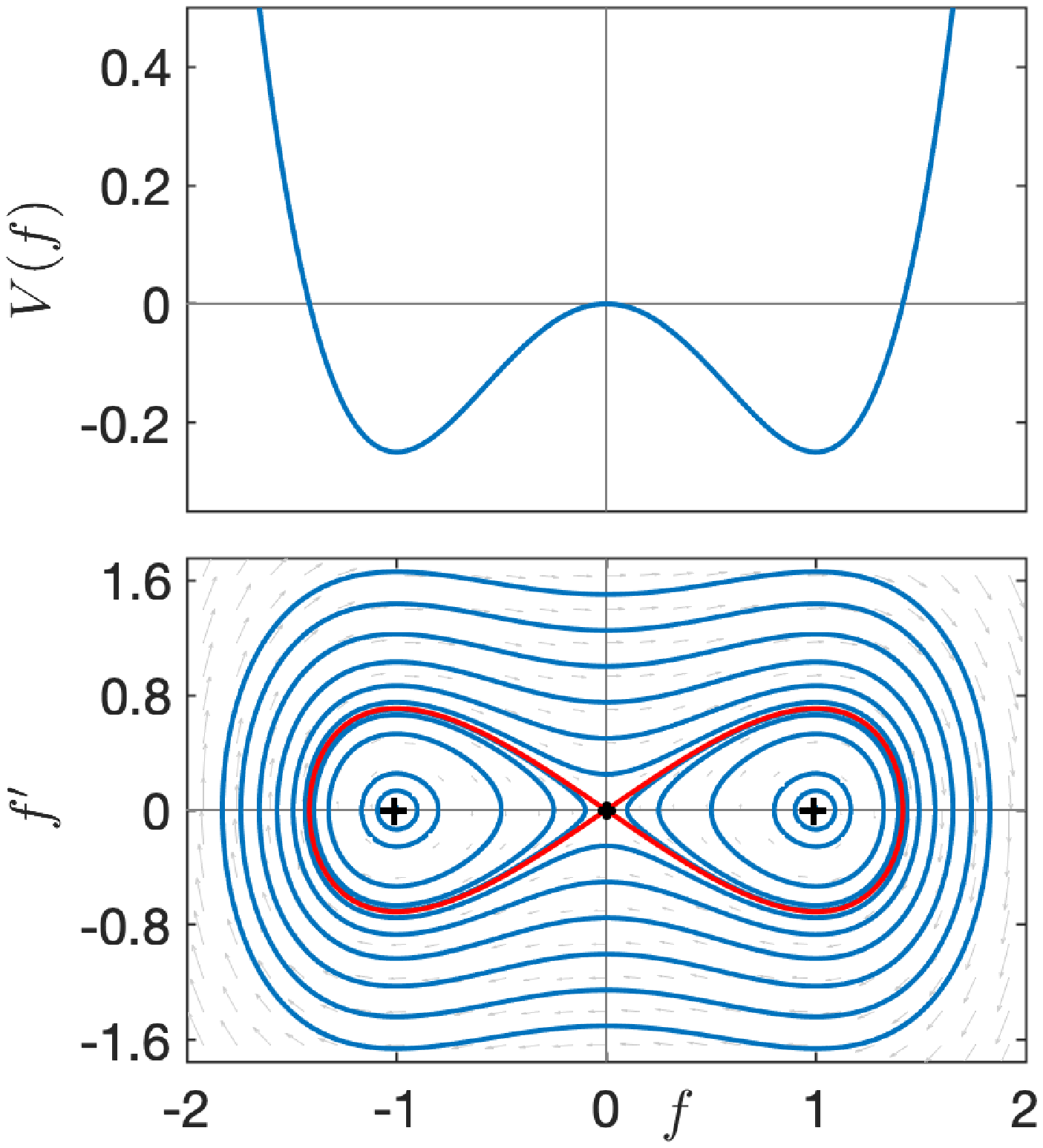}
  \includegraphics[scale=0.470]{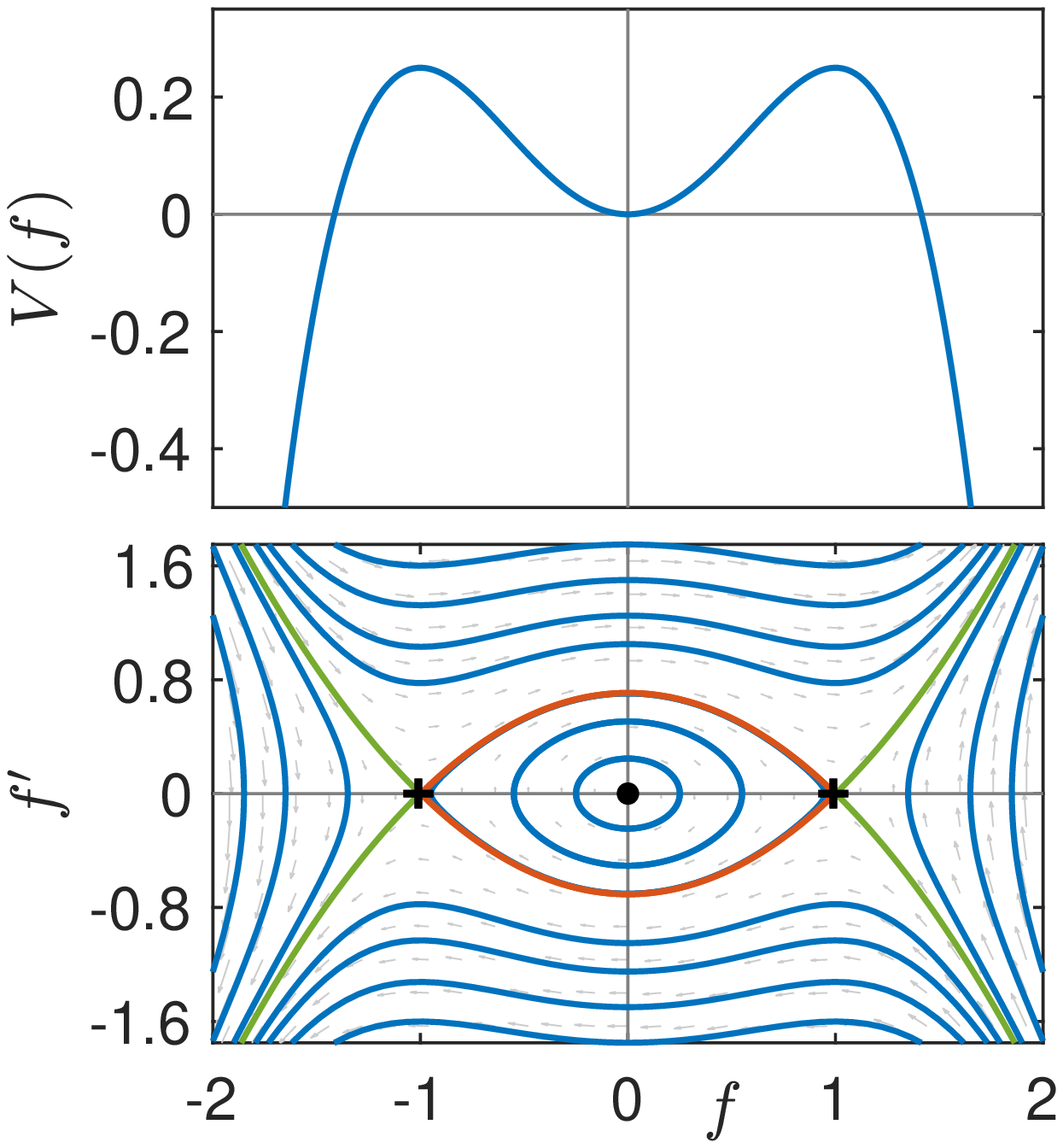}\\
\caption{\label{fig:epsart1}
Left: Typical $(f^{'},f)$ phase plane for $c^2>1$ (in particular, for $c^2-1=1$)  in
the superluminal regime; Right: Typical
$(f^{'},f)$ phase plane for $c^2<1$ (in particular, for
$c^2-1=-1$, i.e., the standing wave case of $c=0$) in
the subluminal regime. The homoclinic and heteroclinic orbits are shown in (red)
color. The corresponding effective potentials associated with the
phase portraits are also shown in the top panel of the figure.}
\end{figure}

As noted above, the relevant solutions and associated existence results
have been previously  presented in~\cite{b21}. Here
we briefly review the corresponding findings by means of first
integral (ODE) considerations.
We examine general traveling wave solutions to~(\ref{eqm}). Defining $y=x-c t$ and $\tau=t$, 
we rewrite~(\ref{eqm}) as
	\begin{equation}
		\label{eq2}
		\left(c^{2}-1\right) \phi_{y y}-2 c \phi_{y \tau}+\phi_{\tau \tau}=\phi-\phi^{3}.
	\end{equation}
We assume that $c\neq1$. We aim to find stationary solutions
to~(\ref{eq2}) in this so-called ``co-traveling'' frame and thus seek
those
in the form $\phi(y, \tau)=f(y)$.
Eq.~(\ref{eq2}) is then reduced to the following standard Duffing equation form:
\begin{equation}\label{eq3}
\left(c^{2}-1\right)f^{''}(y)=f(y)-f(y)^{3},
\end{equation}
Accordingly, the relevant first integral 
\begin{equation}\label{eq4}
\frac{1}{2} f^{'2}(y)+\frac{\frac{1}{2}f(y)^4-f(y)^2}{2 (c^2-1)}=\frac{d}{c^2-1},
\end{equation}
where $d$ is the integration constant.  Letting $V(f)=
\frac{\frac{1}{2}f(y)^4-f(y)^2}{2 (c^2-1)}$ and $\beta=\frac{d}{c^2-1}$, Eq.~(\ref{eq4}) reads
\begin{equation}\label{eq5}
\frac{1}{2} f^{'2}(y)+V(f)=\beta.
\end{equation}
We call stationary solutions $f(y)$ with waves speeds satisfying $c^2<1$ ($c^2>1$) subluminal (superluminal).
Representative phase portraits of subluminal and superluminal solutions are shown in Figure~\ref{fig:epsart1}.
The graphs of $V$ versus $f$ are also shown in the top panel of the
figure. 

$\bullet$ For $c^2>1$ (superluminal), the periodic solutions exist for
$\beta\in(\hat{\beta}_{1}, 0)$ and $\beta>0$. Here
$\hat{\beta}_{1}={\rm min} V(f)$. 
When $\beta\in(\hat{\beta}_{1}, 0)$, the left graph of
Figure~\ref{fig:epsart1} gives rise to two distinct families of
periodic solutions, one for each well of the potential $V$. As the
value of $\beta$ is lowered and approaches the minimum of the well,
the amplitude of the periodic solution shrinks to zero, until it
degenerates to a constant solution.
When $\beta=0$, we retrieve the homoclinic solutions to the problem.
In the phase plane (left panel of Figure~\ref{fig:epsart1}), the solitons arise as homoclinic connections, separating two distinct classes of periodic solutions that exist for $\beta\in(\hat{\beta}, 0)$.  For $\beta>0$, the solutions are bounded and periodic.

$\bullet$ For $c^2<1$ (subluminal),  the periodic solutions exist for
$\beta\in(0, \hat{\beta}_{2})$, where $\hat{\beta_{2}}={\rm max}
V(f)$. Periodic solutions are separated from unbounded solutions by
two heteroclinc orbits, as shown in Figure~\ref{fig:epsart1} (right
panel). As the value of $\beta$ approaches zero, the amplitude of the
periodic solution shrinks to zero, until it degenerates to the
constant vanishing solution. 
 When $\beta=\hat{\beta}_{2}$, the kink solutions appear. The kink solutions arise as heteroclinc connections. When $\beta<0$, the solutions are unbounded.

As stated before, Ref~\cite{b21} has shown three types of elliptic
solutions. We list them here for the purposes of our subsequent stability analysis.

\begin{itemize}
		\item[$\bullet$] The dnoidal waves~\cite{b21}:
	\end{itemize}
\begin{equation}
	\label{eq12}
	f=\beta_{1} {\rm dn}(ly,k),
\end{equation}
	where
	\begin{equation}
		\label{eq13}
		\beta_{1}^2=\frac{2}{2-k^2}, \quad l^2=\frac{\beta_{1}^2}{2(c^2-1)},
	\end{equation}
while the speed $c$ satisfies the condition
\begin{equation}
	\label{eq14}
	|c| \in\left(1, \sqrt{1+\frac{L^{2}}{2 \pi^{2}}}\right),
\end{equation}
	where $L=\frac{2K(k)}{l}$.

\begin{itemize}
		\item[$\bullet$] The cnoidal waves~\cite{b21}:
	\end{itemize}	
	\begin{equation}
		\label{eq18}
		f=\beta_{2} {\rm cn}(ly,k),
	\end{equation}
	where
	\begin{equation}
		\label{eq19}
		\beta_{2}^2=\frac{2k^2}{2k^2-1}, \quad l^2=\frac{\beta_{2}^2-1}{c^2-1},
	\end{equation}
	while the speed $c$ satisfies the condition
	\begin{equation}
		\label{eq20}
		|c| \in\left(1 ,\infty \right),
	\end{equation}
	where $k \in \left(\frac{1}{\sqrt{2}},1\right)  $ and
        $L=\frac{4K(k)}{l}$.	These
        two families are the superluminal ones.
	
	\begin{itemize}
		\item[$\bullet$] The subluminal snoidal waves~\cite{b21}:
	\end{itemize}
	\begin{equation}
		\label{eq15}
		f=\sqrt{2-\beta_{3}^2} {\rm sn}(ly,k),
	\end{equation}
	where
	\begin{equation}
		\label{eq16}
		\beta_{3}^2=\frac{2}{1+k^2}, \quad l^2=\frac{\beta_{3}^2}{2(1-c^2)},
	\end{equation}
	while the speed $c$ satisfies the condition
	\begin{equation}
		\label{eq17}
		|c| \in\left(C_{sb} ,1\right),
	\end{equation}
	where $C_{sb}^2={\rm max}\left\lbrace 0,1-\frac{L^2}{4\pi^2}\right\rbrace  $ and $L=\frac{4K(k)}{l}$.
	
Here $K(k)$ is the complete elliptic integral of the first kind~\cite{lawd} and can be defined as
\begin{align}
    K(k) = \int_0^{\pi/2} \frac{d y}{\sqrt{1-k^2\sin^2(y)}}.
\end{align}

Based on the subluminal snoidal solutions~(\ref{eq15}), it is noted
that if $L>2\pi$, one may consider letting $c\rightarrow{0^{+}}$. In
such case, one could find a real valued periodic standing wave
solutions to~(\ref{eqm}) by letting $c=0$ in~(\ref{eq15});
see also the Remark~3.6 of Ref~\cite{b21} explaining this case.	

\subsection{Stability Analysis}
        
To examine the spectral stability of the elliptic solutions of interest, we consider
\begin{equation}
		\label{eq14b}
	\phi(y, \tau)=f(y)+\epsilon w(y, \tau)+\mathcal{O}\left(\epsilon^{2}\right),
	\end{equation}
where $\epsilon$ is a small parameter. 	
Substituting~(\ref{eq14b}) into equation (\ref{eq2}) and equating terms of order $\epsilon$, we obtain 
\begin{equation}
	\label{eq5b}
	\left(c^{2}-1\right) w_{y y} -2 c w_{y \tau}+w_{\tau \tau}-w+3 f^{2} w=0.
\end{equation}
Using $w_1=w$ and $ w_2=w_{\tau} $, we can write the relevant
linearization partial
differential
equation (PDE) as a first-order system of the form:
\begin{equation}
	\label{eq6}
\frac{\partial}{\partial \tau}\left(\begin{array}{c}
	w_1 \\
	w_2
\end{array}\right)=J \mathcal{L}\left(\begin{array}{c}
	w _1\\
	w_2
\end{array}\right)=J\left(\begin{array}{cc}
	L_{+} & S_{+} \\
	S_{-} & L_{-}
\end{array}\right)\left(\begin{array}{l}
	w_1 \\
	w_2
\end{array}\right),
\end{equation}
	where
	\begin{equation}
		\label{eq7}
		J=\left(\begin{array}{cc}
			0 & 1 \\
			-1 & 0
		\end{array}\right).
	\end{equation}
$L_{-}, L_{+}, S_{+}$and $S_{-}$are defined by
	\begin{equation}
		\label{eq8}
		\begin{aligned}
			&L_{-}=1, \\
			&L_{+}=(c^2-1) \partial_{y y} +3f^2-1 ,\\
			&S_{+}=-2c \partial_{y}, \\
			&S_{-}=0 .
		\end{aligned}
	\end{equation}
Since~(\ref{eq6}) is autonomous in $\tau$,  we separate variables to look for solutions of the form
	\begin{equation}
		\label{eq9}
		\left(\begin{array}{l}
			w_1(y, \tau) \\
			w_2(y, \tau)
		\end{array}\right)=\mathrm{e}^{\lambda \tau}\left(\begin{array}{l}
			W_1(y, \lambda) \\
			W_2(y, \lambda),
		\end{array}\right).
	\end{equation}
Therefore, the spectral problem  is expressed as
	\begin{equation}
		\label{eq11}
		\lambda\left(\begin{array}{l}
			W_1 \\
			W_2
		\end{array}\right)=J \mathcal{L}\left(\begin{array}{l}
			W_1 \\
			W_2        
		\end{array}\right)=J\left(\begin{array}{cc}
			L_{+} & S_{+} \\
			S_{-} & L_{-}
		\end{array}\right)\left(\begin{array}{l}
			W_1 \\
			W_2
		\end{array}\right).
	\end{equation}
We define the spectrum $\sigma(J \mathcal{L})$ of the operator $J \mathcal{L}$
\begin{equation}
\sigma_{J\mathcal{L}}=\left\{\lambda \in \mathbb{C} : \sup _{x \in \mathbb{R}}\left(\left|W_{1}(x)\right|,\left|W_{2}(x)\right|\right)<\infty\right\}.
\end{equation}
\\
\\
\textbf{Definition 1} \emph{If there exists $\lambda$ with
  $\operatorname{Re}(\lambda)>0$, then the stationary elliptic
  solution is called spectrally unstable.  It is called
  \textbf{modulationally unstable} if the unstable spectral band with
  $\operatorname{Re}(\lambda)>0$ intersects the origin in the
  $\lambda$-plane. This definition of modulationally unstable is in
  line, e.g., with~\cite{b23,b26,b27}.}
\\

In this section, we study the spectral instability and modulation
instability using Hill's method originally proposed in~\cite{b36}.  To
apply Hill's method,  we need Fourier expansions for $f^2$
of~(\ref{eq11}). Using the complex Fourier
series expansion, we obtain
\begin{equation}\label{f2}
f^{2}(y)=\sum_{n=-\infty}^{\infty} Q_{n} e^{i 2 n \pi y / L},
\end{equation}
where $Q_{n}$ are Fourier coefficients.

Since $f^2(y)$ is a periodic function, following from Floquet's theorem, we decompose the eigenfunction components $W_{1}$ and $W_{2}$ in a Fourier-Floquet form 

\begin{equation}\label{f1}
W_{1}(y)=e^{i \mu y} \sum_{n=-\infty}^{\infty} W_{1,n} e^{i n 2\pi y / PL} \quad \text { and } \quad W_{2}(y)=e^{i \mu y} \sum_{n=-\infty}^{\infty} W_{2,n} e^{i n 2\pi y / PL},
\end{equation}
where $\mu\in[-\frac{\pi}{PL}, \frac{\pi}{PL}]$, and $W_{1,n}$ and $W_{2,n}$ are Fourier coefficients.

Substituting~(\ref{f1}) and~(\ref{f2}) into~(\ref{eq11}) and equating
Fourier coefficients , we write the
equations for $W_{1,n}$ and $W_{2,n}$ as a coupled bi-infinite system
	\begin{equation}
		\label{eq22}
		\begin{aligned}
			 &W_{2,n}=\lambda W_{1,n}, \\
			&\left(1+(c^2-1)(\mu+\frac{2\pi n}{PL})^2\right) W_{1,n}+2ci\left(\mu+\frac{2\pi n}{PL}\right) W_{2,n}\\
			&-3 \sum_{m=-\infty}^{\infty} Q_{(n-m) / P} \delta_{P \mid (n-m)}W_{1,n}=\lambda W_{2,n},
		\end{aligned}
	\end{equation}
Here $Q_{\frac{n-m}{P}}=0$ if $\frac{n-m}{P} \notin \mathbb{Z}$, and
$\delta_{P \mid(n-m)}$ is 1 if P divides $(n-m)$, and 0 otherwise.
By choosing a finite number of Fourier modes, the exact bi-infinite
system of Eq.~(\ref{eq22}) is truncated. We explicitly compute
approximations to the spectral elements by searching the eigenvalues
of the truncation of~(\ref{eq22}). 

\begin{figure}
\hspace{-1cm}\includegraphics[scale=0.600]{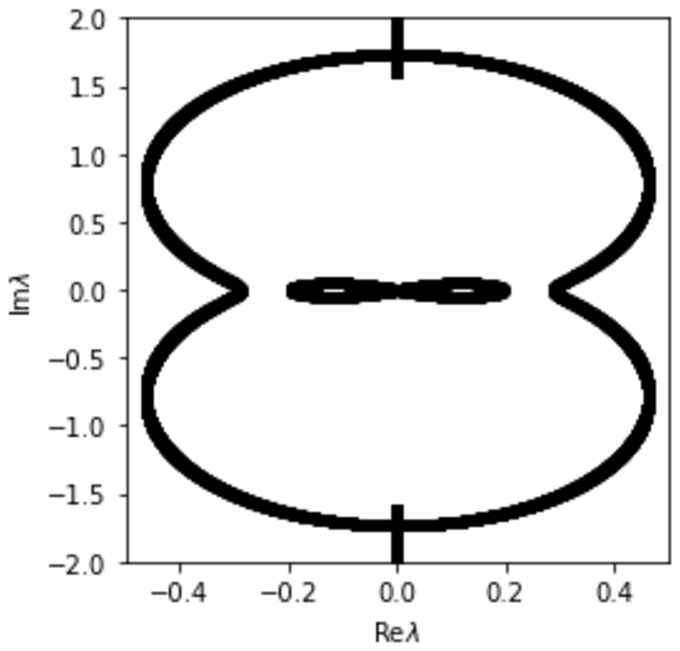} \includegraphics[scale=0.600]{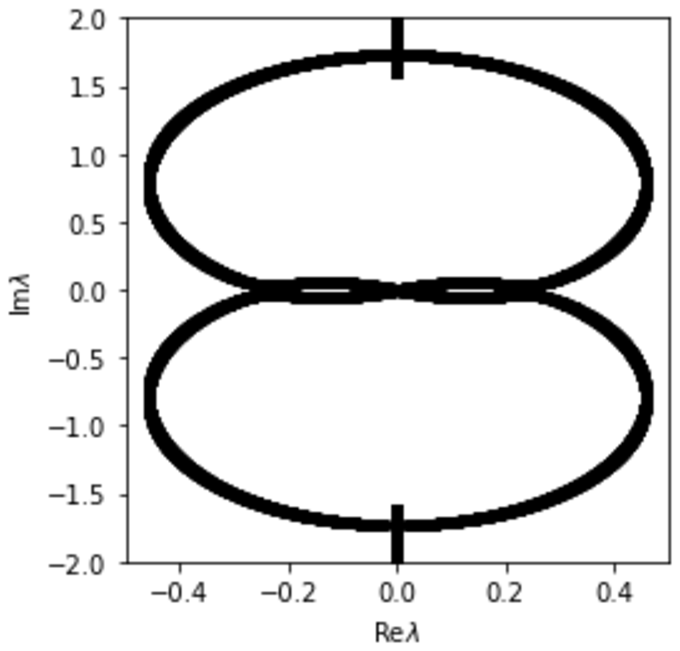}  \includegraphics[scale=0.600]{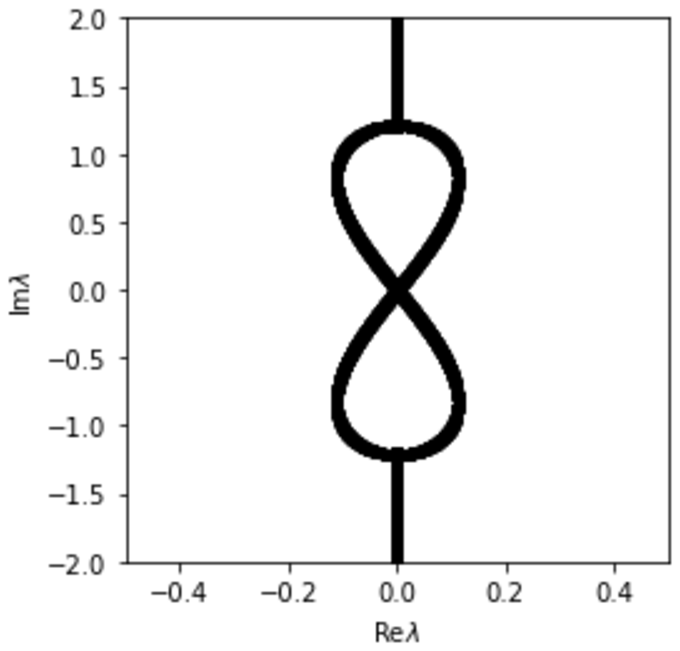}\\
\caption{\label{fig:epsart3}
  Left: The stability spectrum (i.e., the imaginary part of $\lambda$
  on the vertical vs. the real part of $\lambda$ on the horizontal) for the dnoidal solutions in
the superluminal regime for $c=1.9$ and $k=0.999$.
Middle:  The stability spectrum for dnoidal solutions in
the superluminal regime for $c=1.9$ and $k=0.9989$.
Right: The stability spectrum for dnoidal solutions in
the superluminal regime for $c=1.9$ and
$k=0.5$.
Here, and in the figures that follow, we have ensured that the panels
illustrate representative cases. 
}
\end{figure}

\begin{figure}
\hspace{-1cm}\includegraphics[scale=0.600]{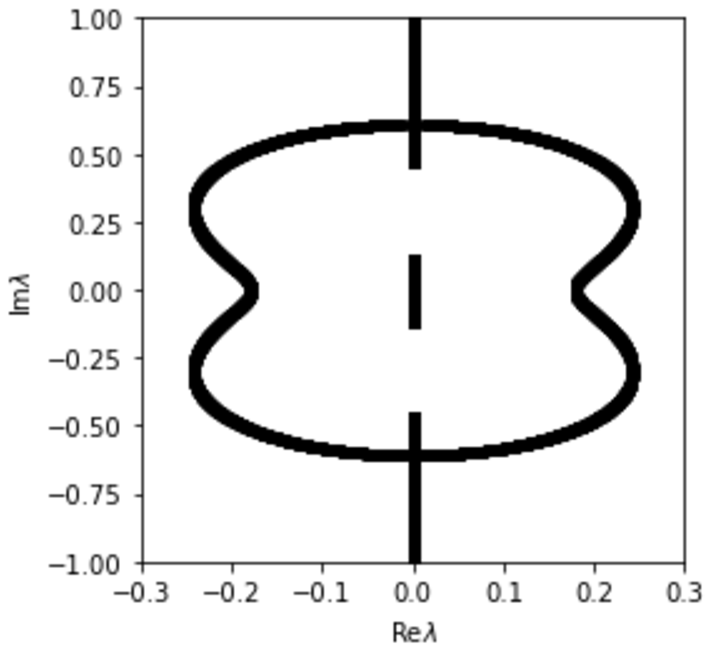} \includegraphics[scale=0.600]{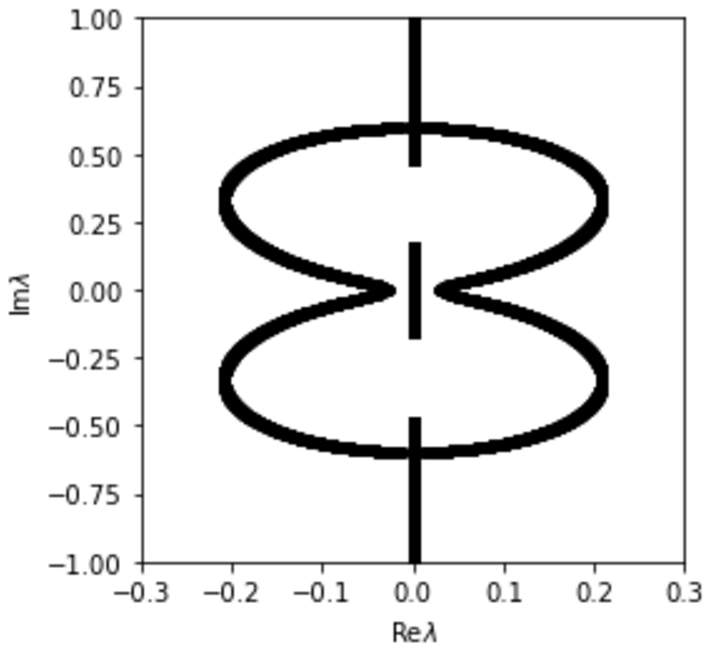}  \includegraphics[scale=0.600]{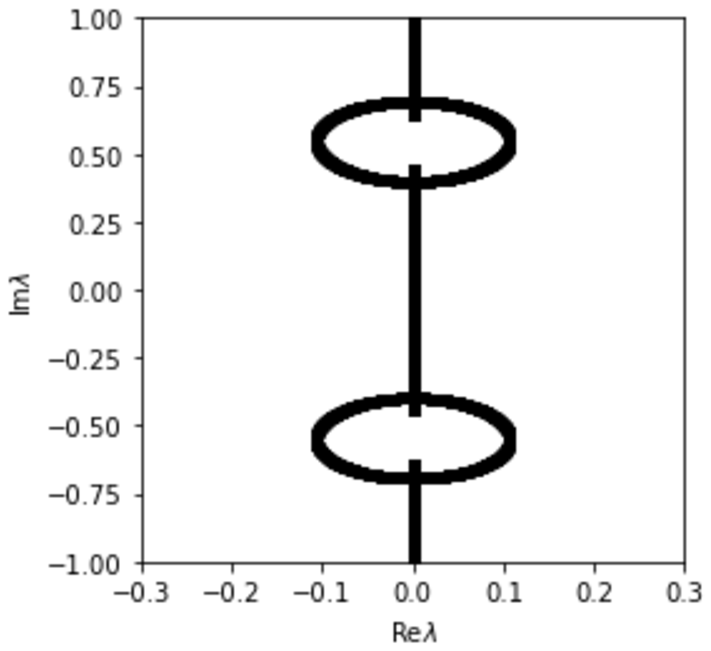}\\
\caption{\label{fig:epsart4}
  Left: The stability spectrum (i.e., the imaginary part of $\lambda$
  on the vertical vs. the real part of $\lambda$ on the horizontal)
  for cnoidal solutions in
the superluminal regime for $c=1.1$ and $k=0.99$.
Middle: The stability spectrum for cnoidal solutions in
the superluminal regime for $c=1.1$ and $k=0.9825$.
Right: The stability spectrum for cnoidal solutions in
the superluminal regime for $c=1.1$ and $k=0.9$.}
\end{figure}

For the dnoidal waves, the closure of $\sigma_{\mathcal{JL}}$ not on
the imaginary axis consists of either an infinity symbol inset inside
an ellipse-like curve, see Figure~\ref{fig:epsart3} (left; see also
the middle panel) or a figure 8, see Figure~\ref{fig:epsart3} (right).
For the cnoidal waves, the situation is different. In particular,
as
shown in  Figure~\ref{fig:epsart4} (left; see also the middle panel), 
the closure of $\sigma_{\mathcal{JL}}$ not on the imaginary axis may
consist of
an ellipse-like curve surrounding the origin. {\it However}, it  is
also
possible for different parameters to have
two ellipse-like curves in the upper- and lower-half plane, as is
shown
in the right panel of Figure~\ref{fig:epsart4}.

\begin{figure}
\hspace{-1cm}\includegraphics[scale=0.600]{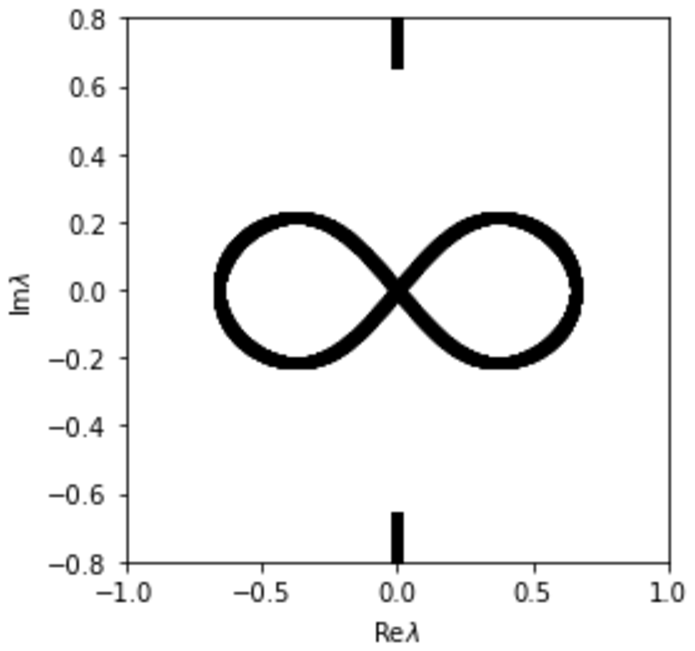} \includegraphics[scale=0.600]{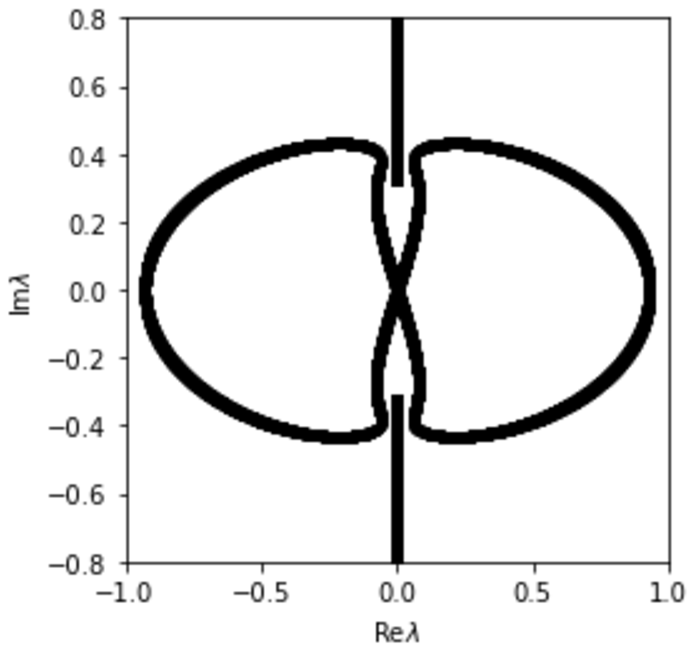}  \includegraphics[scale=0.600]{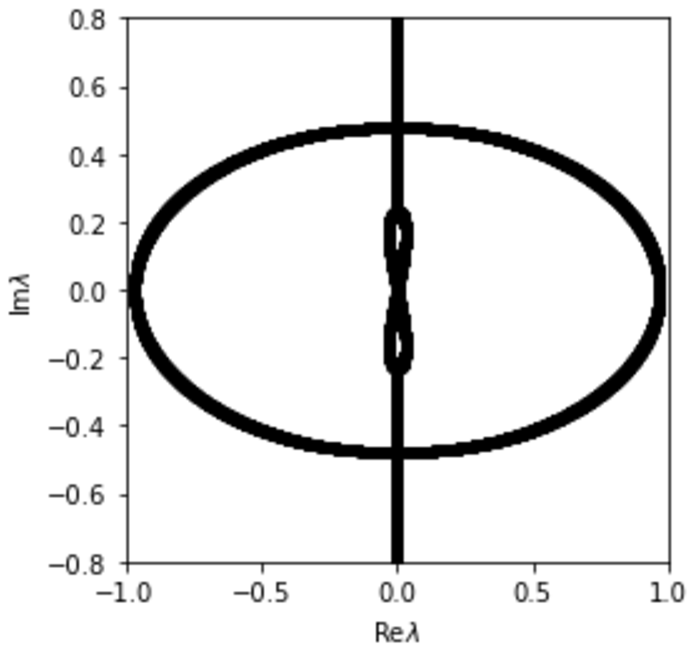}\\
\caption{\label{fig:epsart5}
  Left: The stability spectrum (in a similar representation as the
  previous figures) for snoidal solutions  in
the subluminal regime for $c=0.5$ and $k=0.5$; 
Middle: The stability spectrum for snoidal solutions  in
the subluminal regime for $c=0.5$ and $k=0.22$;
Right: The stability spectrum for snoidal solutions  in
the subluminal regime for $c=0.5$ and $k=0.15$.}
\end{figure}

Turning to the subluminal case, for snoidal waves (with $c=0.5$), the closure
of $\sigma_{\mathcal{JL}}$ not on the imaginary axis consists of
either an infinity symbol, as observed in Figure~\ref{fig:epsart5}
(left), or  a figure 8
inset inside an ellipse-like curve, see
Figure~\ref{fig:epsart5} (right).
The middle panel of the figure shows how the former case gradually
deforms
into the latter. From Figure~\ref{fig:epsart5}, we find that although
the unstable spectral bands vary with different $k$, such unstable
spectral bands always intersect with
the origin in the $\lambda$-plane, which implies that the snoidal waves are modulationally unstable.

Based on the above observation, the modulational
instability/stability results
can be summarized as follows.
\\
\\
\textbf{Modulational stability/instability results:} \emph{In the
  superluminal regime,  the dnoidal solutions are spectrally unstable
  and the spectrum forms a figure eight intersecting at the origin,
  which implies, based on the above Definition 1,  that they are
  subject to modulational instability (as shown in
  Figure~\ref{fig:epsart3}).
  On the other hand, the cnoidal solutions are, typically,
  modulationally stable
  based on the above definition, since the spectrum near the origin is
  represented
  by the vertical bands along the purely imaginary axis (see
  Figure~\ref{fig:epsart4}).
  Finally, in the subluminal regime, the snoidal solutions are
  modulationally unstable  since the spectrum forms a figure eight
  intersecting at the origin (as seen in Figure~\ref{fig:epsart5})}.

\begin{figure}
\hspace{-0.55cm}\includegraphics[scale=0.400]{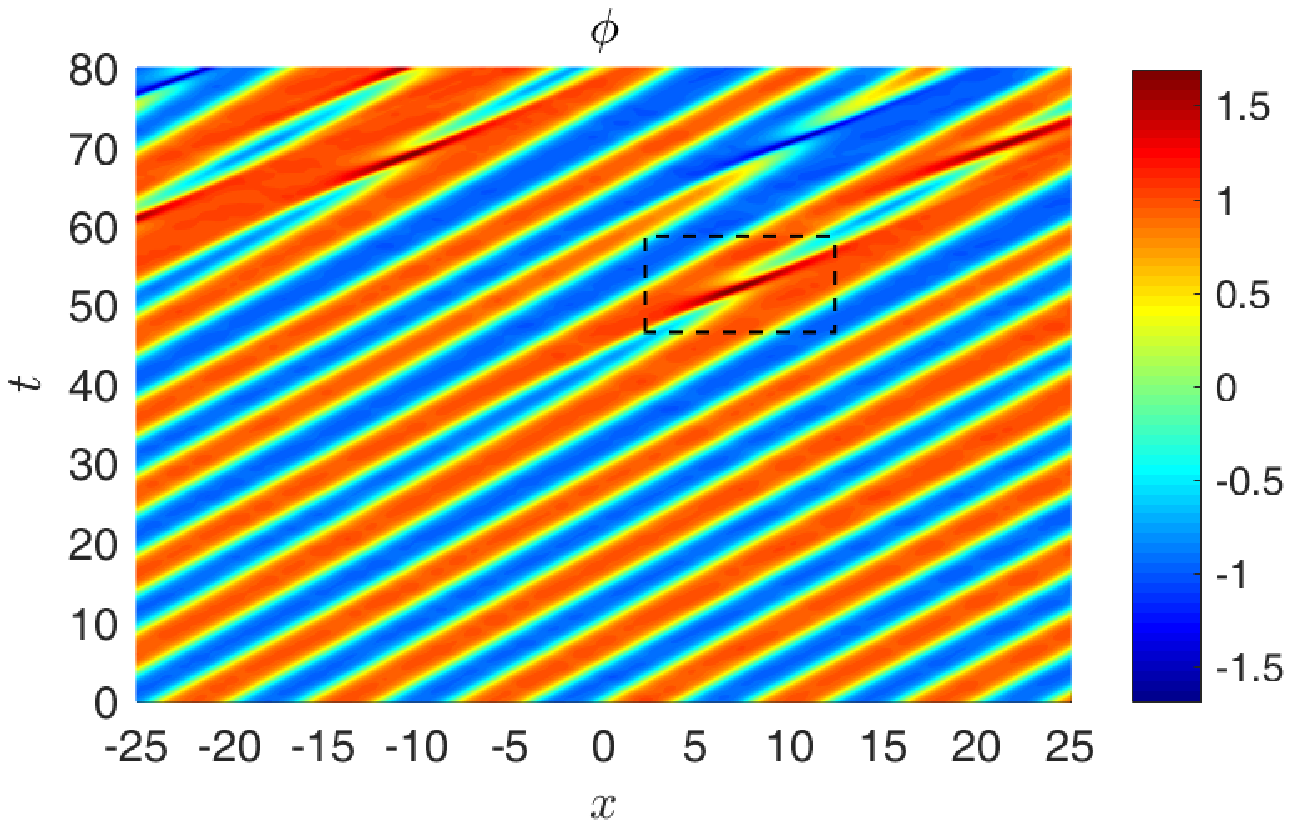}\quad \    \includegraphics[scale=0.400]{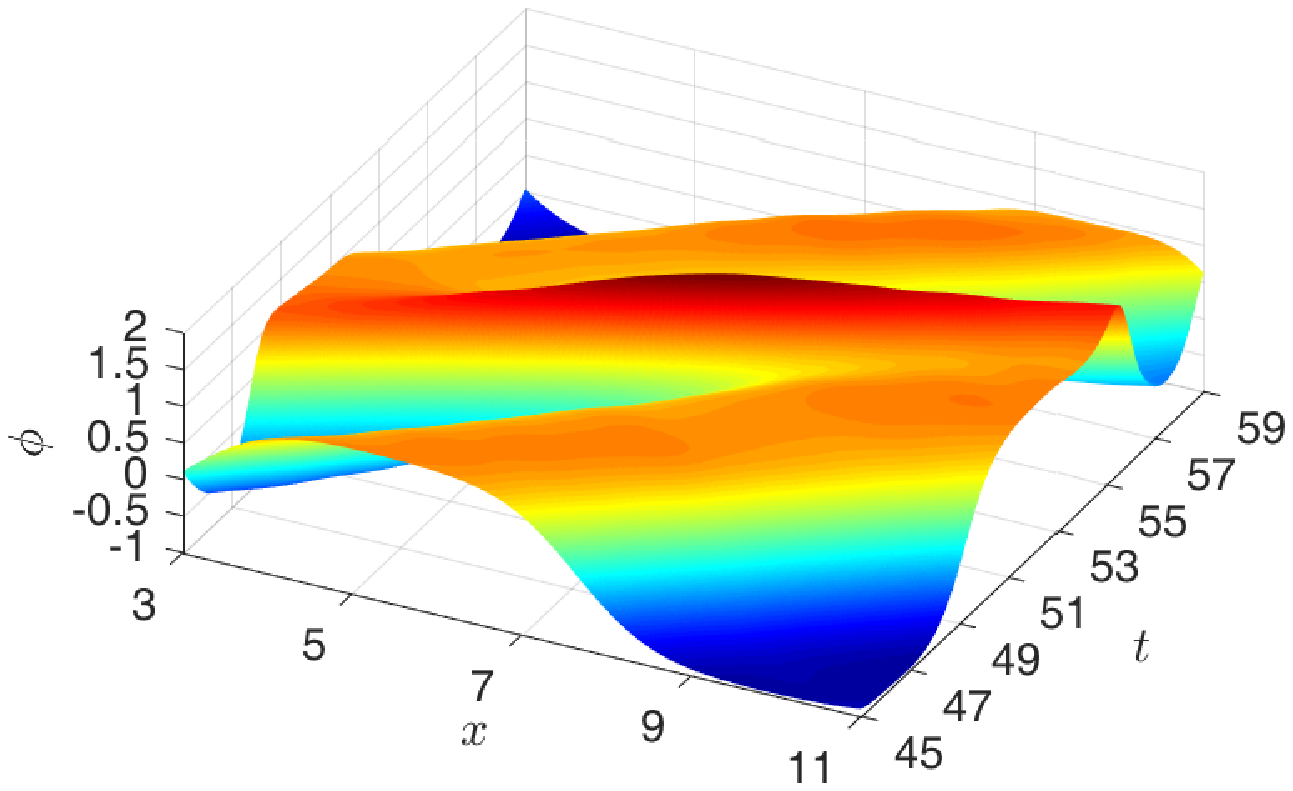}\ \ \includegraphics[scale=0.4500]{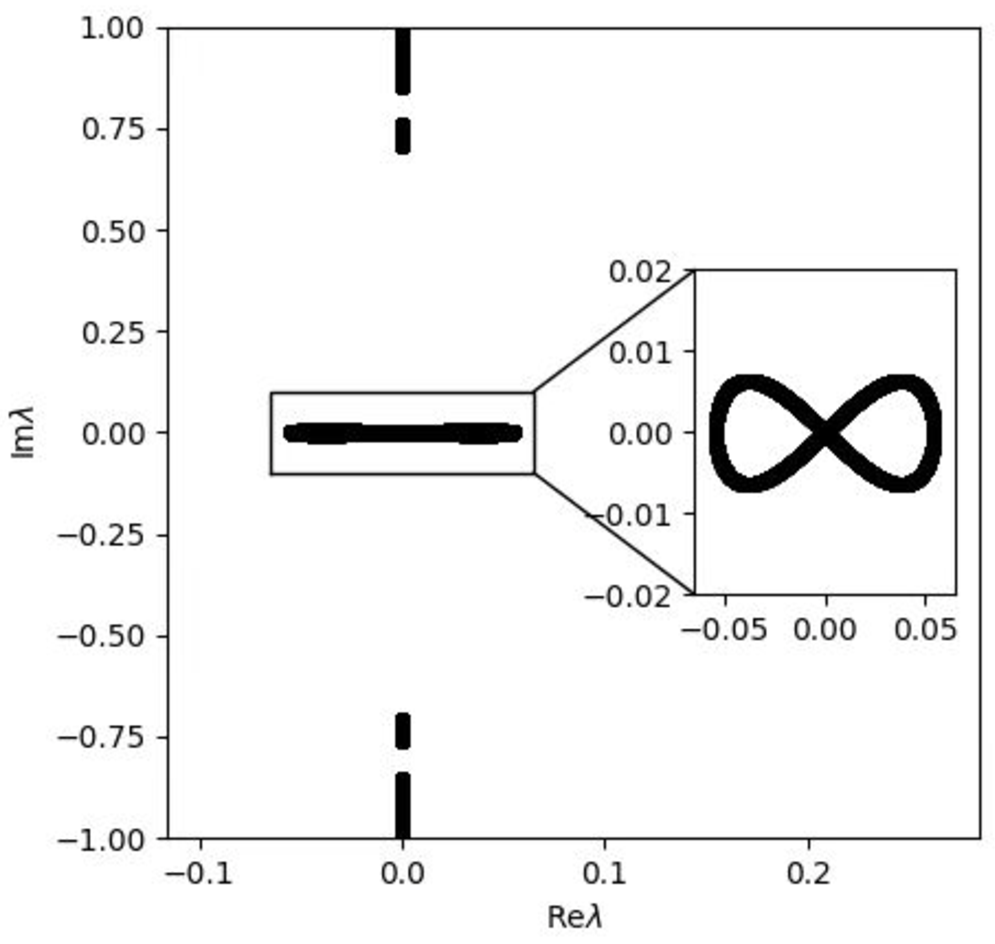}\\
\caption{\label{fig:epsart14}
Numerical excitation of localized structures on the  subluminal snoidal solution background. The initial condition is a snoidal wave perturbed by $5 \%$ random noise with $k=0.93$ and $c=0.8$.
The amplitude evolution (left); the enlarged 3D plot on the right
illustrates the spatio-temporally localized wave pattern  highlighted
on the left  by a surrounding box (middle); the corresponding stability spectrum for snoidal solutions in
the subluminal regime with $k=0.93$ and $c=0.8$ (right).}
\end{figure}

\begin{figure}
\hspace{-1cm}\includegraphics[scale=0.400]{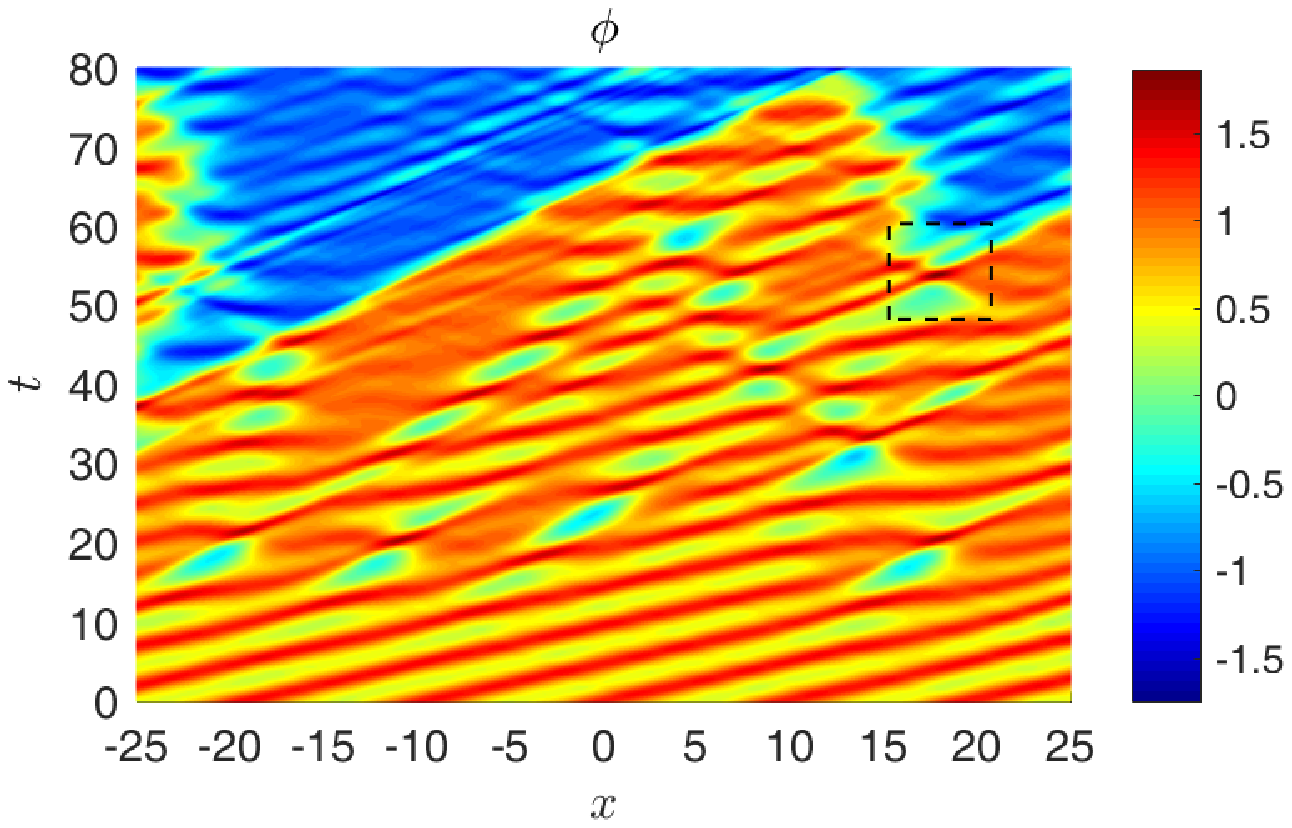}\quad \ \ \  \includegraphics[scale=0.400]{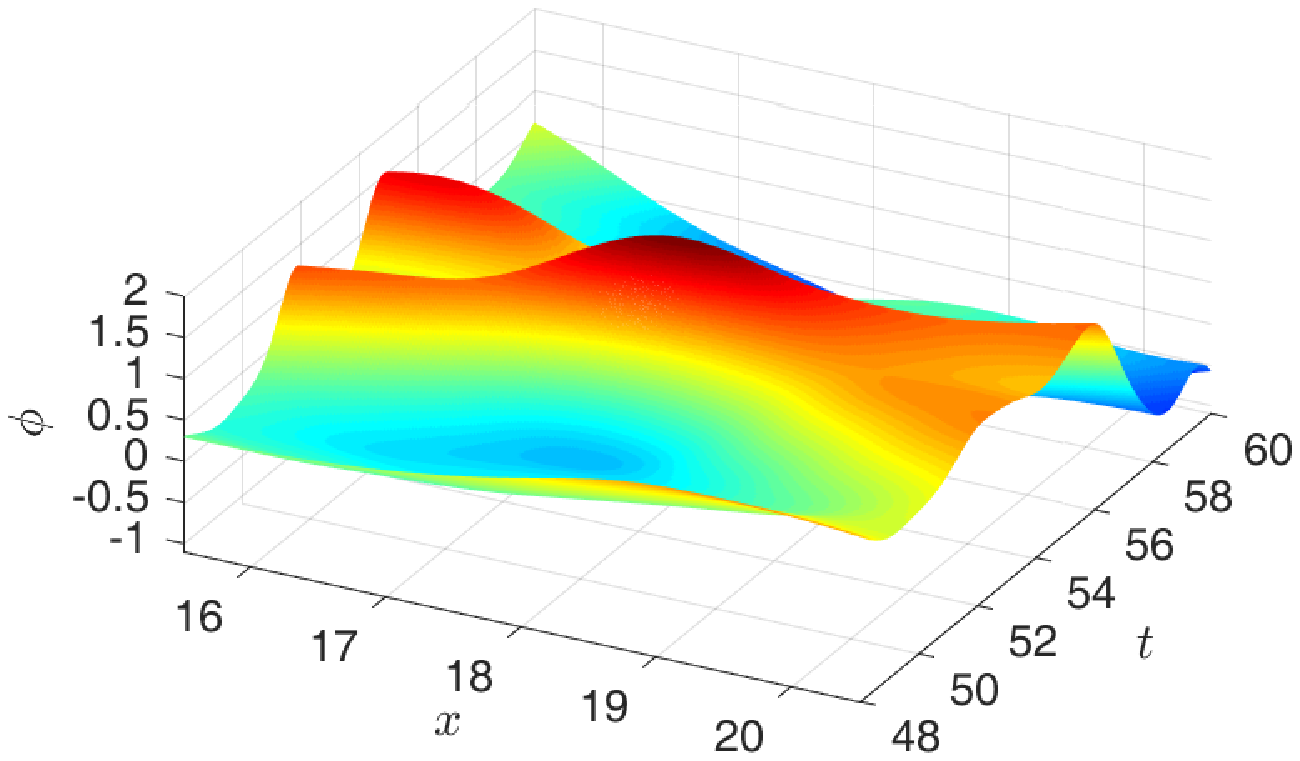}\ \ \ \includegraphics[scale=0.5500]{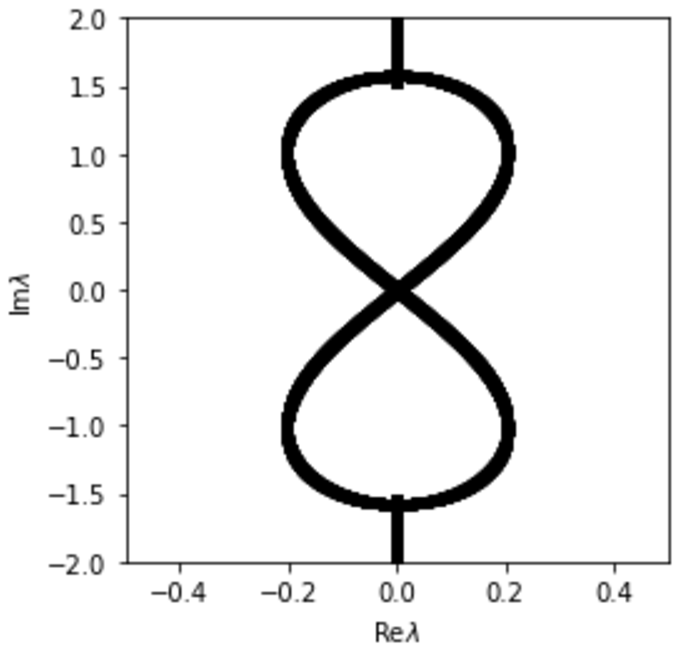}\\
\caption{\label{fig:epsart15}
Numerical excitation of localized structures on the dnoidal
waveform in
the superluminal regime. The initial condition is a dnoidal wave perturbed by $5 \%$
random noise with $k=0.96$ and $c=2$. The amplitude evolution (left);
once again the zoomed-in evolution of the box on the left is shown on
the middle panel; the corresponding stability spectrum for dnoidal solutions in
the superluminal regime with $k=0.96$ and $c=2$ is shown on the right panel.}
\end{figure}

\begin{figure}
\hspace{-1cm}\includegraphics[scale=0.3500]{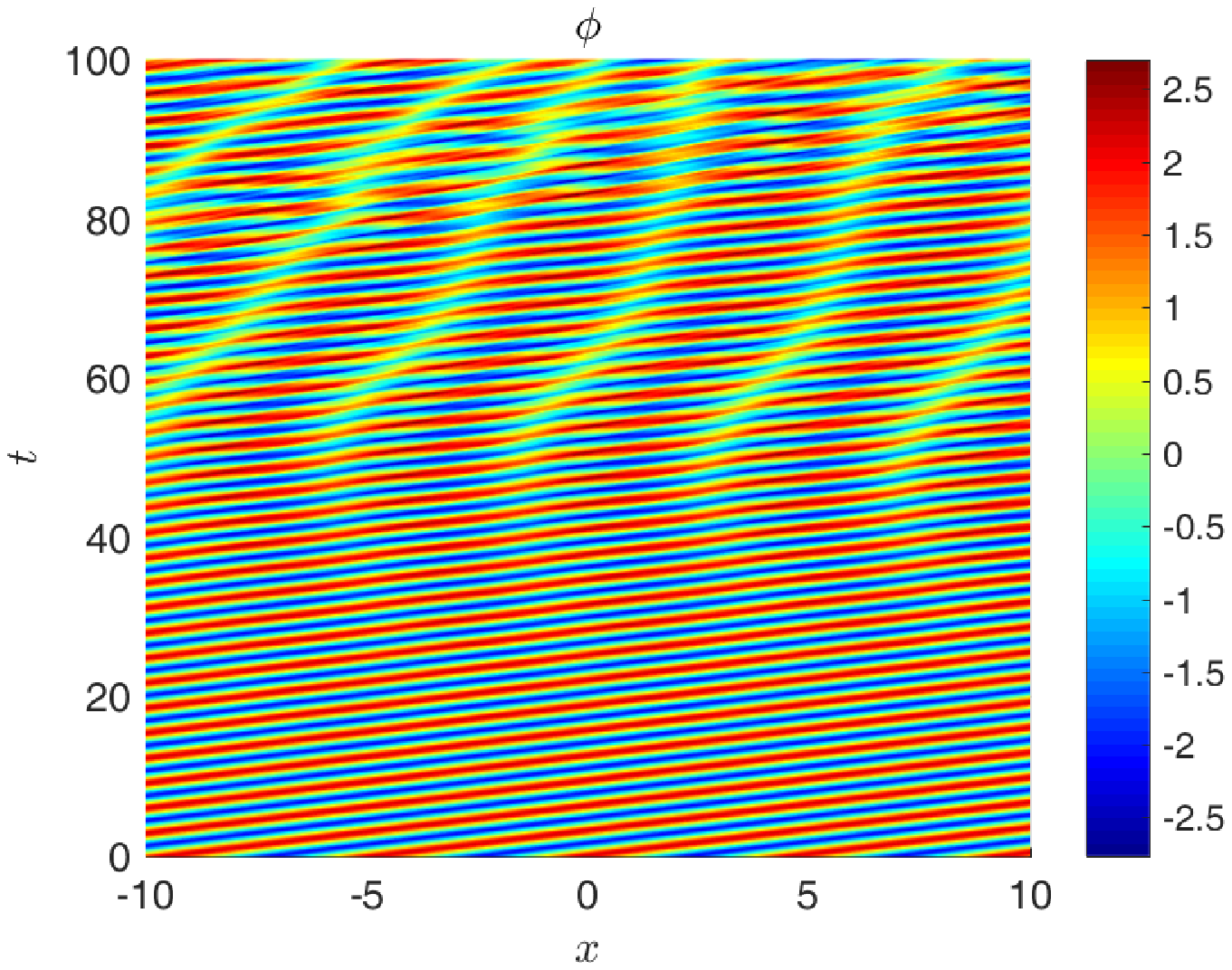}\quad\ \ \ \includegraphics[scale=0.550]{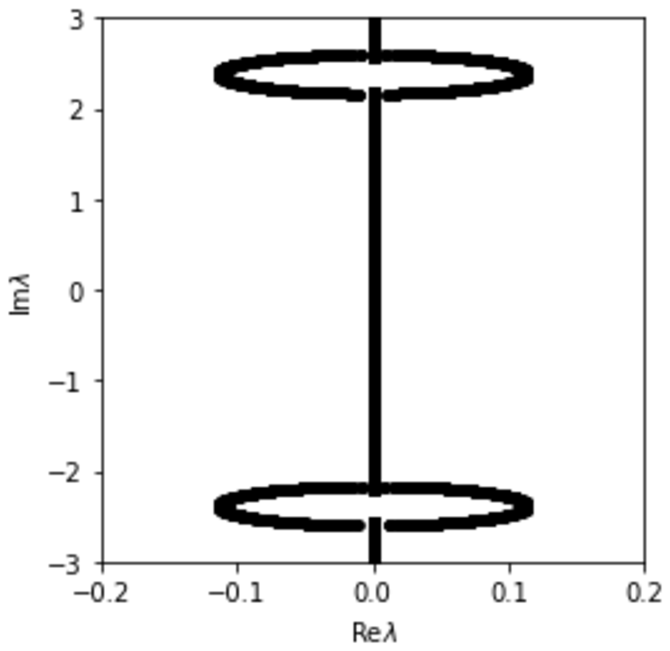}\\
\caption{\label{fig:epsart15s}
Numerical excitation of the instability of the cnoidal
waveform in
the superluminal regime. The initial condition is a cnoidal wave perturbed by $5 \%$
random noise with $k=0.8$ and $c=1.5$ (left); the corresponding stability spectrum for cnoidal solutions in
the superluminal regime with $k=0.8$ and $c=1.5$ (right).} 
\end{figure}

\textbf{Definition 2} \emph{Let $\phi$ be  a periodic travelling wave
  of the $\phi^{4}$ equation with the period $L$ and
  $\widetilde{\phi}$ be another solution to the $\phi^{4}$
  equation. One can say that $\widetilde{\phi}$ is a 
  localized wave (LW) on top of the spatio-temporally (ST) periodic
background (LWST) $\phi$  if $\widetilde{\phi}$ is different from the orbit $\left\{\phi\left(x-x_0\right)\right\}_{x_0 \in[0, L]}$ for $t \in \mathbb{R}$ and it satisfies}


\begin{equation}
\inf _{x_0 \in[0, L]} \sup _{x \in \mathbb{R}}\left|\tilde{\phi}(x, t)-\phi\left(x-x_0\right)\right| \rightarrow 0 \quad \text { as } \quad t \rightarrow \pm \infty.
\end{equation}

It is generally recognized that modulational instability is among the
mechanisms which generate spatio-temporally localized structures. 
Therefore, we expect that the LWSTs can be generated on the
modulationally unstable backgrounds.  It is worthwhile to note that
the work of~\cite{b27} has identified rogue wave on top of a
periodic background following a definition similar to Definition 2. 
Since the $\phi^4$ model is not integrable, such potential rogue wave
solutions
are not available in analytical form. In light of that, our approach here
will be to attempt to identify possible LWSTs numerically, i.e.,
through
direct numerical simulations of modulationally unstable waveforms.

More concretely, we use the split-step Fourier spectral method to
investigate the
possible waveforms generated from modulationally unstable backgrounds. For this purpose, we simulate the evolution of the elliptic solutions taken as the initial condition, perturbed by  random noise of relative strength  5\%.  Specifically, the perturbation is added as  $\phi(x,0)+0.05 f(x)$, where $f(x)$ is the white noise perturbation. 
A prototypical example of this form is shown 
in Figure~\ref{fig:epsart14} (left and middle) where the numerical excitation of
LWs on top of the snoidal solution background
can be detected. Besides, the modulationally unstable for snoidal solution in
the subluminal regime with $k=0.93$ and $c=0.8$ is shown in Figure~\ref{fig:epsart14} (right).
Indeed, notice that the original structure propagates at constant
slope in the spatio-temporal contour plot, featuring, for a while, a
perfect traveling
wave (despite the, imperceptible to the eye, random noise of 5\% of
the original solution imposed upon the initial condition).
However, eventually, the modulational instability sets in
producing at distinct instances (such as the one boxed in the figure) of
localized excitations on top of the spatio-temporally periodic
background. These are states that we refer to as LW on ST backgrounds.

A far more dramatic event can be observed in the case of the dnoidal
wave background in Figure~\ref{fig:epsart15} (left and middle).
Figure~\ref{fig:epsart15} (right) shows that the  dnoidal solution with $k=0.96$ and $c=2$ is modulationally unstable. 
Firstly, it is worth
noting
that this solution is considerably more unstable as the modulational
instability manifests itself faster.
Secondly, it is relevant to note that in this case, when
they first appear,  the  LWSTs do not represent bumps but rather {\it dips} on top
of the modulationally unstable background.
Nevertheless, bumps may also occur, as in the boxed interval of the
evolution,
zoomed into on the right panel of the figure.
Finally, the deepening of
the above-mentioned dips results in the nucleation of pairs of ``kink-like''
structures
that subsequently lead to collisions and resulting complex dynamics
(leading to an evolution in the vicinity of the other well of the
double well potential). It is interesting to compare/contrast this
behavior to that associated with the modulationally stable
case of the cnoidal waves, shown in Figure~\ref{fig:epsart15s} (right).
Importantly, in the latter, the instability is manifested as well,
however, as can be seen by the spatially localized ``threads''
resulting
from the instability and persisting over the time evolution,
the waveforms emerging do not feature
spatio-temporal localization (as in the dnoidal case of the previous
two figures), as shown in Figure~\ref{fig:epsart15s} (left).




\noindent\textbf{\large 3. Stability and instability of elliptic solutions with respect to subharmonic perturbations}\\\hspace*{\parindent}

Finally, in this section, we consider the spectral stability of
elliptic solutions with respect to a $P$-subharmonic perturbation,
which is a perturbation of integer multiple of $P$ times the period of
the solution. We start by illustrating the transition process from a
spectrally unstable state to a spectrally stable state with respect to
a fixed $c$ as $k$ decreases. Such a transition implies that dnoidal
and cnoidal solutions are (or, more precisely,  can become)
spectrally stable for relevant parameter
values with respect to $P$-subharmonic perturbations.  
Secondly, we show why snoidal solutions remain unstable with respect to all subharmonic perturbations.

As shown before, since~(\ref{eq11}) is a spectral problem with
periodic coefficients, the Floquet-Bloch decomposition results in the
following form of eigenfunctions
(written now compactly as):
\begin{equation}
\left(\begin{array}{l}
W_{1}(y) \\
W_{2}(y)
\end{array}\right)=e^{i \mu y}\left(\begin{array}{c}
\hat{W_{1}}(y) \\
\hat{W_{2}}(y)
\end{array}\right), \quad \hat{W_{1}}(y+T(k))=\hat{W_{1}}(y), \quad \hat{W_{2}}(y+T(k))=\hat{W_{2}}(y),
\end{equation}
with $\mu \in[0,2 \pi / T(k))$. Here $T(k)=\frac{4K(k)}{l}$ for sn and
cn solutions,
and $T(k)=\frac{2K(k)}{l}$ for dn solutions. 
For $P$-subharmonic perturbations,
\begin{equation}
\mu=l \frac{2 \pi}{P T(k)}, \quad l=0, \ldots, P-1.
\end{equation}

\begin{figure}
\hspace{-1cm}\includegraphics[scale=0.5500]{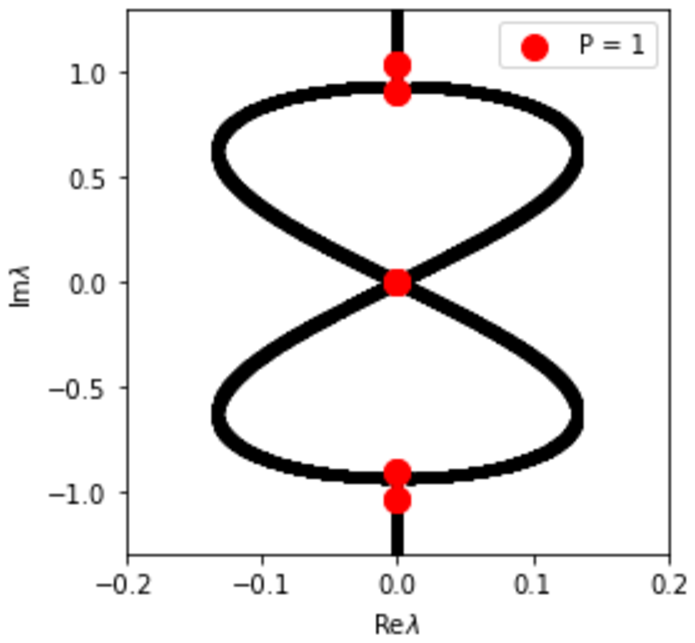}\quad \includegraphics[scale=0.5500]{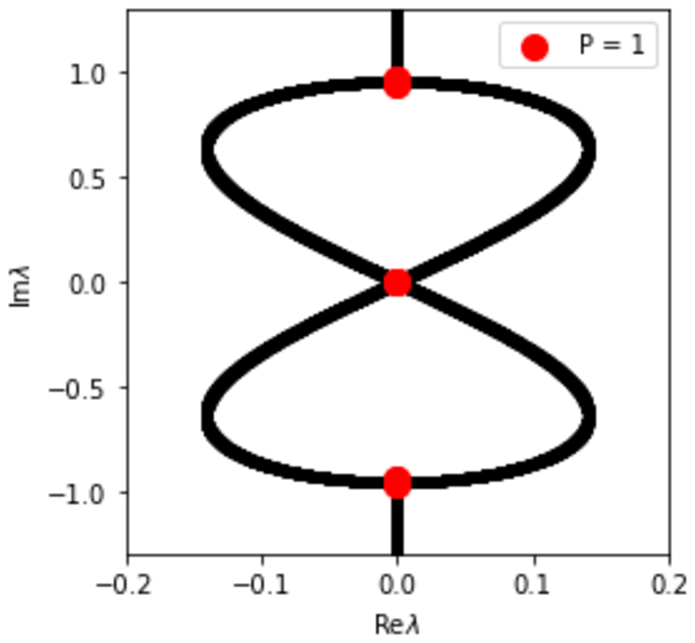}\quad  \includegraphics[scale=0.5500]{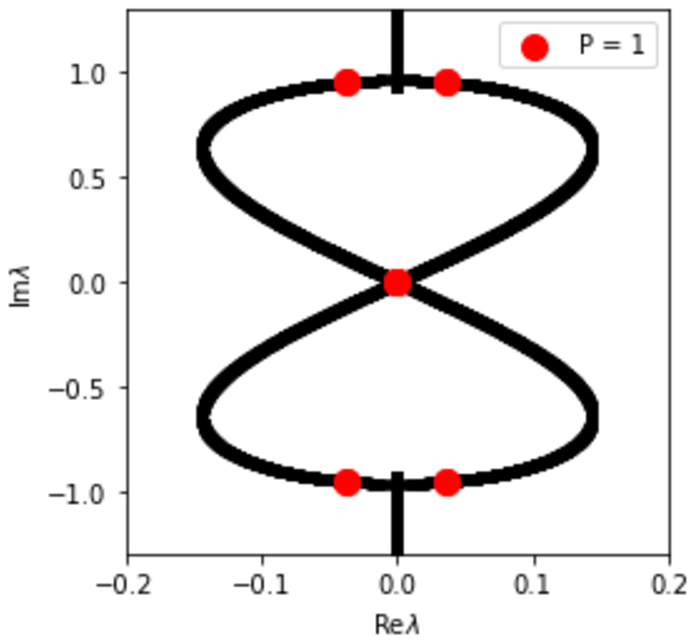}\\
\caption{\label{fig:epsart6}
Left: The stability spectrum  (i.e., the imaginary part of $\lambda$
on the vertical vs. the real part of $\lambda$ on the horizontal)
for dnoidal solutions in
the superluminal regime for $c=1.5$ and
$k=0.93$. Middle: The stability spectrum for dnoidal solutions in
the superluminal regime for
$c=1.5$ and $k=0.936$.  Right: The stability spectrum for dnoidal
solutions in
the superluminal regime for $c=1.5$ and $k=0.938$. While the black curves represent
the full spectrum, the important feature to detect
here and in the following two figures is the $P=1$ spectrum
associated with red dots, denoting the co-periodic spectrum of the
solution of interest.}
\end{figure}

\begin{figure}
\hspace{-1cm}\includegraphics[scale=0.5500]{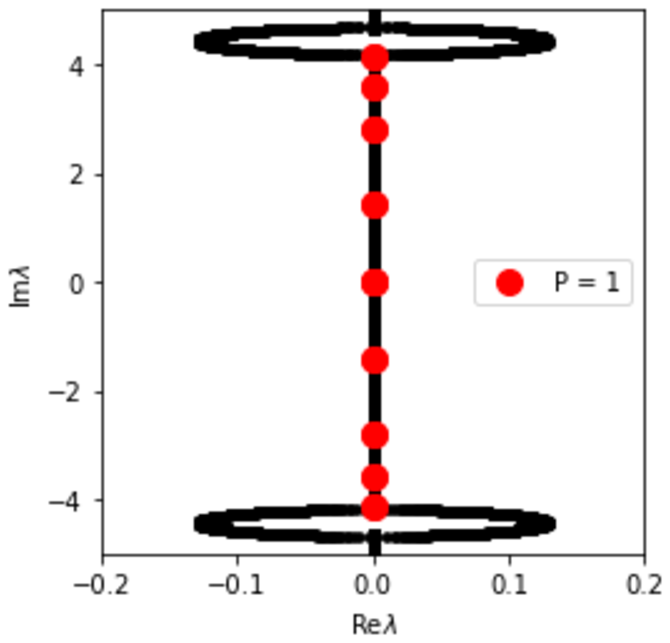}\quad \includegraphics[scale=0.5500]{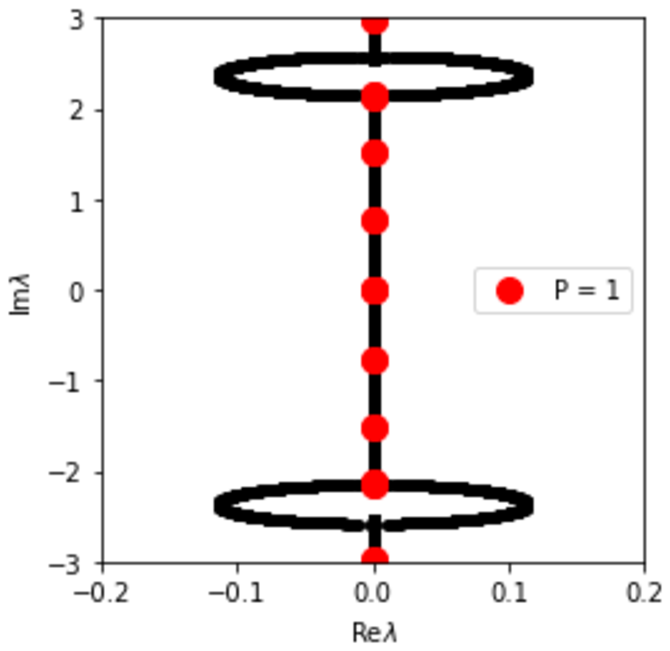}\quad  \includegraphics[scale=0.5500]{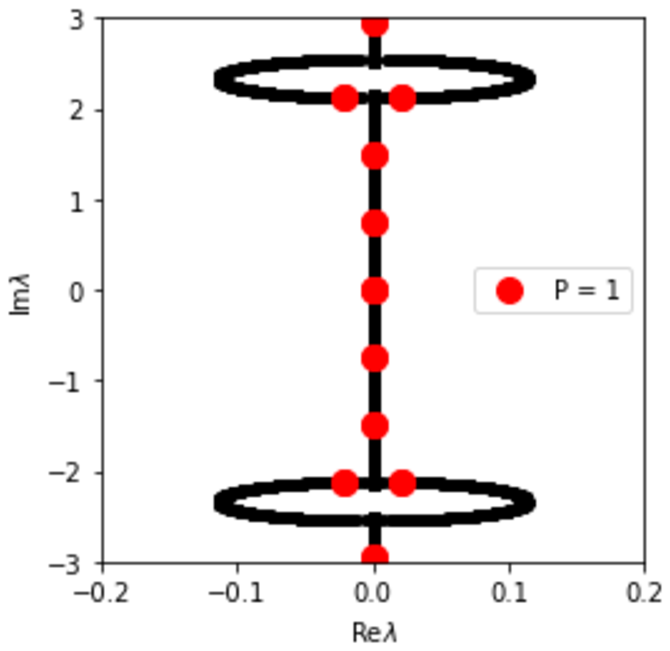}\\
\caption{\label{fig:epsart7}
Left: The stability spectrum (in a similar format as in the above
figure) for
the cnoidal solutions in
the superluminal regime for $c=1.5$ and $k=0.74$. Middle: The stability spectrum for cnoidal solutions in
the superluminal regime for $c=1.5$ and $k=0.801$.  Right: The stability spectrum for cnoidal solutions in
the superluminal regime for $c=1.5$ and $k=0.803$.}
\end{figure}

\begin{figure}
\hspace{-1cm}\includegraphics[scale=0.5500]{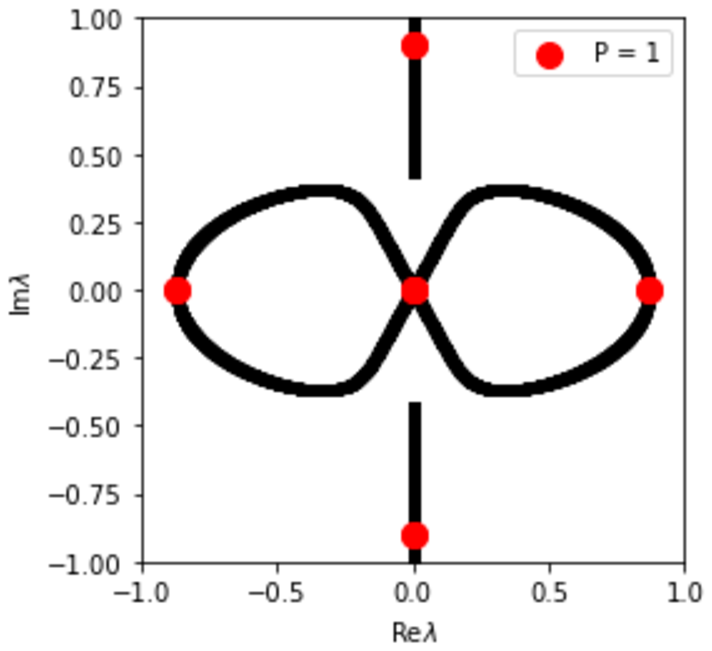}\quad \includegraphics[scale=0.5500]{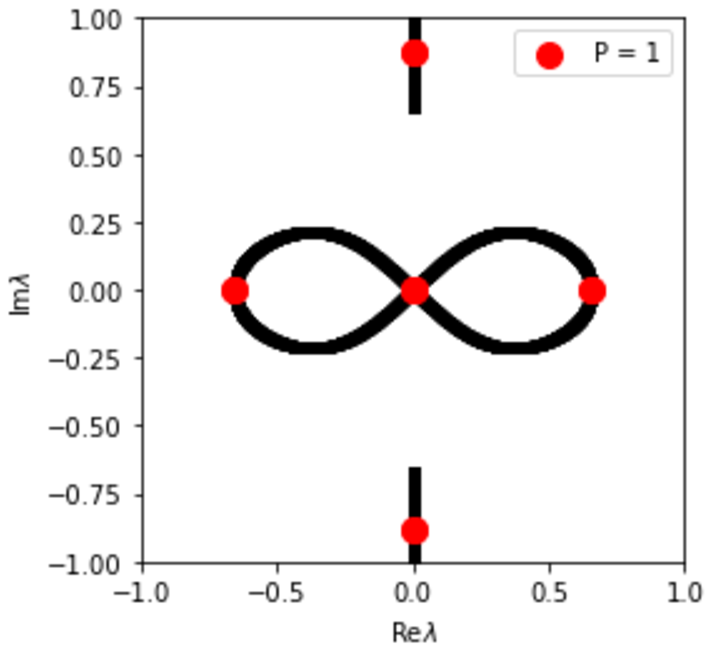}\quad  \includegraphics[scale=0.5500]{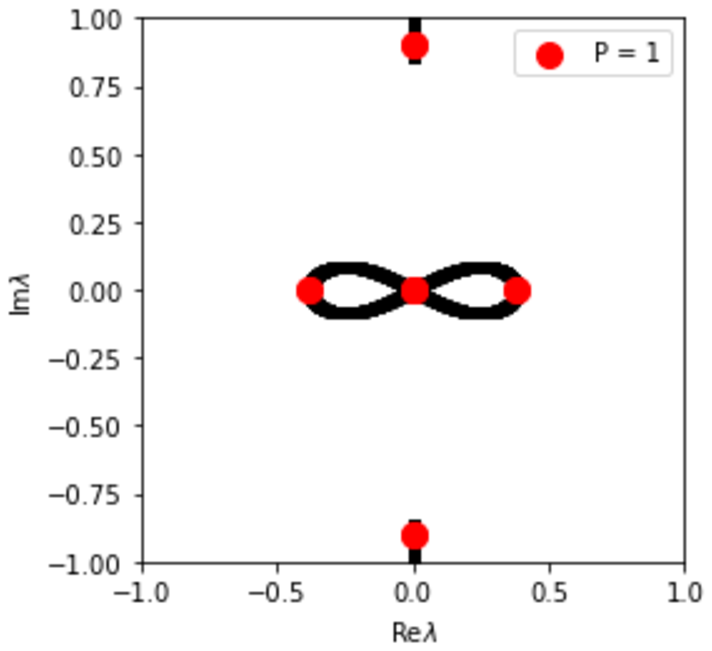}\\
\caption{\label{fig:epsart8}
Left: The stability spectrum (once again in a similar format) for snoidal solutions  in
the subluminal regime  for $c=0.5$ and $k=0.3$. Middle: The stability spectrum for snoidal solutions in
the subluminal regime for $c=0.5$ and $k=0.5$.  Right: The stability spectrum for snoidal solutions  in
the subluminal regime for $c=0.5$ and $k=0.7$.}
\end{figure}

Figures~\ref{fig:epsart6} and~\ref{fig:epsart7} show how  a solution
which is spectrally stable with respect to subharmonic perturbations
loses stability as $k$ is varied. We begin by using Hill's method to
compute the point spectrum for a single subharmonic perturbation. For
dnoidal solutions, we note that two eigenvalues gradually approach
and eventually collide at the intersection of the top of the figure 8
spectrum and the imaginary axis.
Spectral instability occurs when two  imaginary eigenvalues
collide along the imaginary axis, leading to a Hamiltonian Hopf bifurcation and
enter the right
and left half planes along the figure 8, as shown in
Figure~\ref{fig:epsart6}. The same process occurs for cnoidal
solutions, as shown in Figure~\ref{fig:epsart7},
although, as discussed above, this no longer takes place along a
figure 8,
but rather along the two ellipse-shaped curves which occur away from
the origin of the spectral plane. Specifically, for cnoidal solutions, two eigenvalues gradually approach
and eventually collide at the intersection of the ellipse-shaped
spectrum and the imaginary axis, leading to a Hamiltonian Hopf bifurcation and
enter the right
and left half planes along the ellipse-shaped curves, as shown in Figure~\ref{fig:epsart7}.
Since we referred to the Hamiltonian-Hopf bifurcation taking place
in this context, it is interesting to point out that in the $\phi^4$
problem,
similarly to the sine-Gordon case, such bifurcations are {\it entirely
  absent} from the linearization around {\it static} solutions. In the
latter case, the spectral problem is self-adjoint and can thus only
feature real or imaginary eigenvalues. It is only in the case of
traveling
solutions that such non-self-adjoint, speed-dependent terms arise in
the linearization problem of Eqs.~(\ref{eq6})-(\ref{eq8}) and hence
such
Hamiltonian-Hopf bifurcations and complex eigenvalue quartets
are possible in this setting.
From Figure~\ref{fig:epsart8}, we note that there are two eigenvalues
(corresponding to $P=1$) always located on the intersections of the
figure 8 and the real axis, which means that snoidal solutions are
(for all parameter values) spectrally unstable with respect to
co-periodic perturbations, contrary to what was the case for the
previous two families, but in line with the results of~\cite{b21}.

\begin{figure}
\hspace{-1cm}\includegraphics[scale=0.5500]{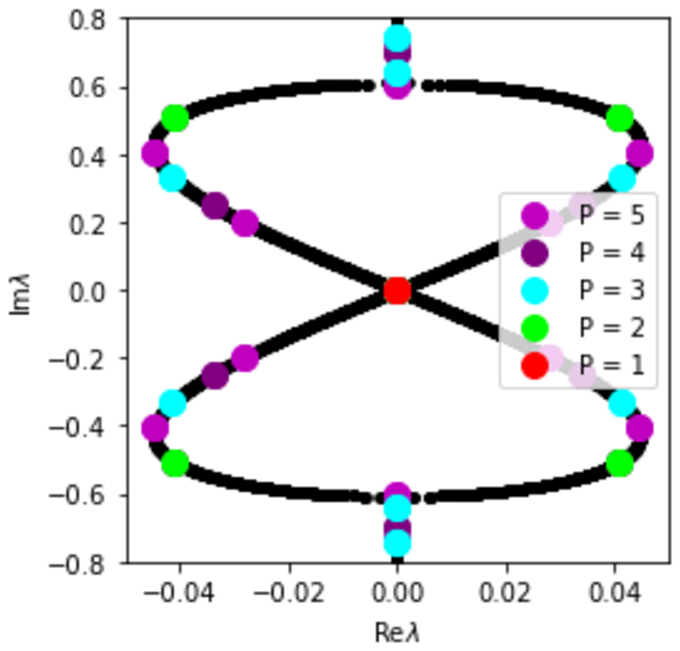}\quad \includegraphics[scale=0.5500]{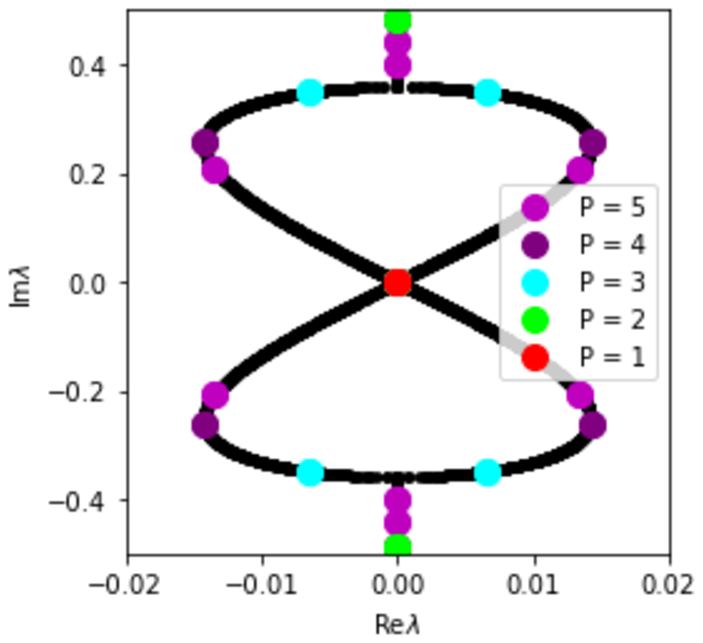} \quad\includegraphics[scale=0.5500]{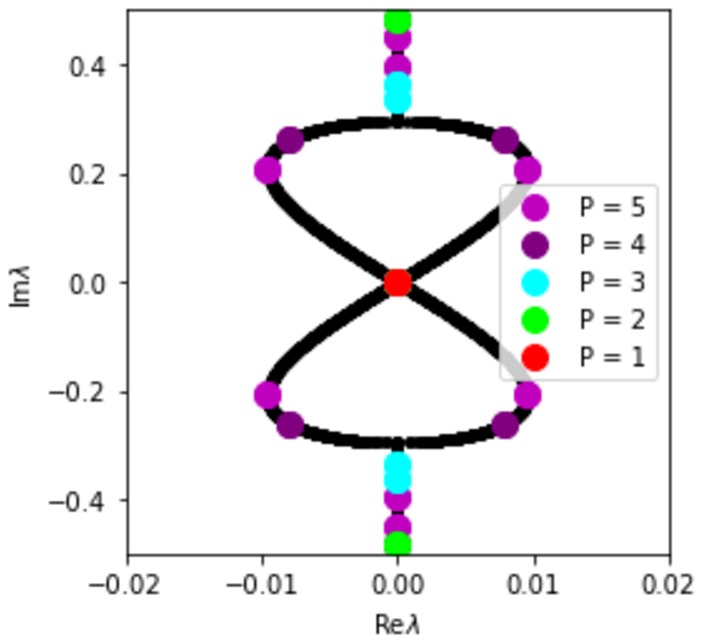}\\
\caption{\label{fig:epsart9}
Left: The stability spectrum for dnoidal solutions  in
the superluminal regime for $c=1.5$ and
$k=0.8$. Middle: The stability spectrum for dnoidal solutions  in
the superluminal regime for
$c=1.5$ and $k=0.65$.  Right: The stability spectrum for dnoidal
solutions  in
the superluminal regime for $c=1.5$ and $k=0.6$. Here and in the following two
figures,
in addition to the
co-periodic
spectrum (red dots), the additional colors (green, blue, purple,
magenta)
represent the spectrum to subharmonic perturbations of higher $P$; see
also the text for details.}
\end{figure}

\begin{figure}
\hspace{-1cm}\includegraphics[scale=0.5500]{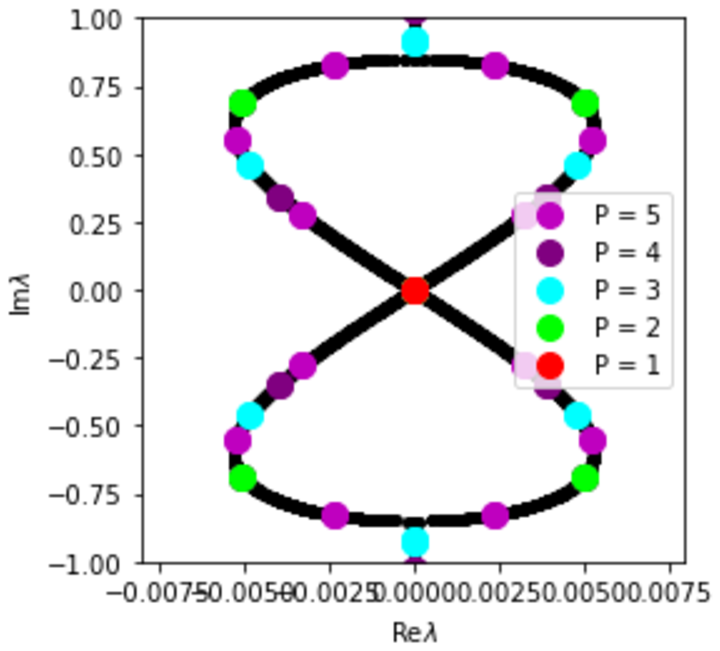}\quad \includegraphics[scale=0.5500]{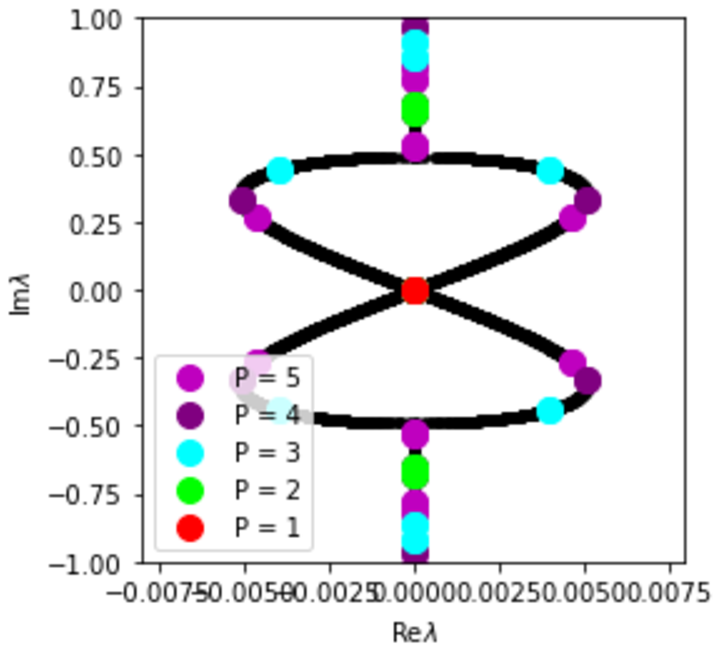} \quad\includegraphics[scale=0.5500]{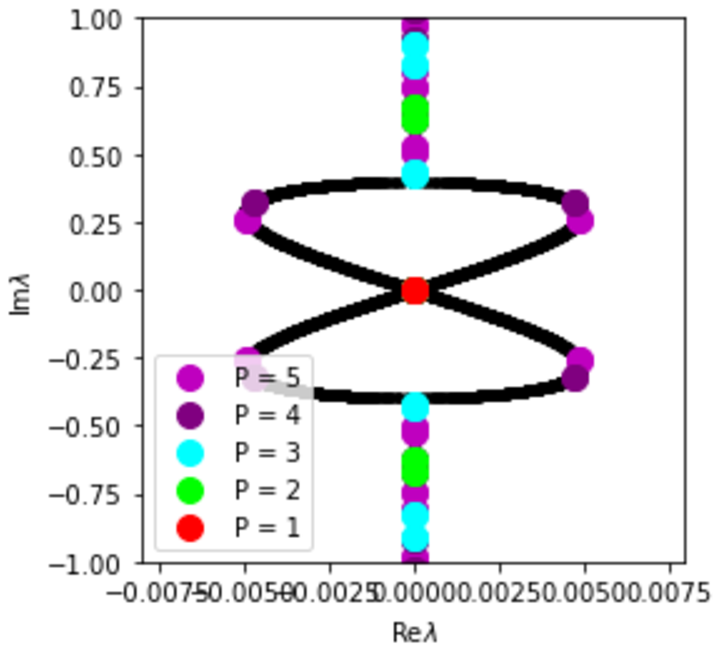}\\
\caption{\label{fig:epsart10}
Left: The stability spectrum for dnoidal solutions in
the superluminal regime for $c=5$ and $k=0.5$. Middle: The stability spectrum for dnoidal solutions in
the superluminal regime for $c=3$ and $k=0.5$.  Right: The stability spectrum for dnoidal solutions in
the superluminal regime for $c=2.5$ and $k=0.5$.}
\end{figure}

\begin{figure}
\hspace{-1cm}\includegraphics[scale=0.800]{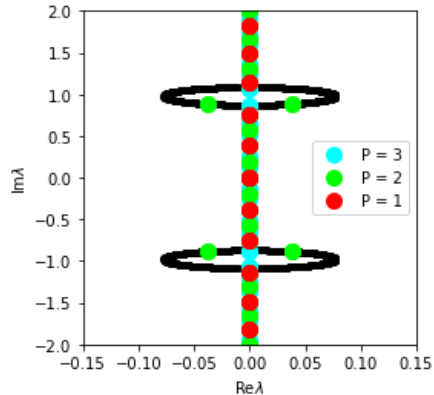}\\
\caption{\label{fig:epsart11}
The stability spectrum for cnoidal solutions in
the superluminal regime for $c=1.1$ and $k=0.8$.}
\end{figure}

Since the transition between spectral stability and instability can be
obtained
in this context through a Hamiltonian Hopf bifurcation, we show some stability results for cnoidal and dnoidal waves in the superluminal regime. 

$\bullet$  For dnoidal solutions, for a fixed $c$, by decreasing $k$,
we conclude that the dnoidal solutions are spectrally stable with
respect to larger $P$-subharmonic perturbations. From
Figure~\ref{fig:epsart9}, we note that when $c=1.5$ and $k=0.8$, the
dnoidal solutions are spectrally stable with respect to co-periodic
perturbations but unstable with respect to other subharmonic perturbations.
When $c=1.5$ and $k=0.65$, the dnoidal solutions are spectrally stable
with respect to 1- and 2-subharmonic perturbations but unstable with
respect to other (higher-$P$) perturbations. When $c=1.5$ and $k=0.6$,
the dn solutions are spectrally stable with respect to 1-, 2- and
3-subharmonic perturbations
but unstable with respect to higher $P$ perturbations. I.e., as $k$ is
decreased,
apparently higher-$P$ perturbations are progressively stabilized.
For dnoidal solutions, for a fixed $k$, by decreasing $c$, we conclude
that the dnoidal solutions are spectrally stable with respect to
larger $P$-subharmonic perturbations, as shown in
Figure~\ref{fig:epsart10}.
We note that when $c=5$ and $k=0.5$,  the
dnoidal solutions are spectrally stable with respect to co-periodic
perturbations but unstable with respect to other subharmonic perturbations, as shown in Figure~\ref{fig:epsart10} (left).
When $c=3$ and $k=0.5$, the dnoidal solutions are spectrally stable
with respect to 1- and 2-subharmonic perturbations but unstable with
respect to other (higher-$P$) perturbations, as shown in Figure~\ref{fig:epsart10} (middle). When $c=2.5$ and $k=0.5$, the dn solutions are spectrally stable with respect to 1-, 2- and
3-subharmonic perturbations
but unstable with respect to higher $P$ perturbations, as shown in Figure~\ref{fig:epsart10} (right). 
 I.e., also decreasing $c$, for fixed $k$,
also
stabilizes higher-$P$ subharmonic perturbations.

$\bullet$ For cnoidal solutions,  we find that
 there are
 always subharmonic perturbations that are unstable, even though
 there may exist both higher and lower values of $P$ for which
 the subharmonic perturbations may be stable.
For example, when $c=1.1$ and $k=0.8$, we find that the cnoidal solutions are spectrally stable with 1- and 3-subharmonic perturbations but not with respect to 2-subharmonic perturbations, as shown in Figure~\ref{fig:epsart11}.
\\
\\
\noindent\textbf{\large 4. Conclusions/Future Work}\\\hspace*{\parindent}

In this work, we have revisited the existence of traveling periodic
wave
patterns in the prototypical non-integrable Klein-Gordon model in the
form
of the $\phi^4$ equation. Our particular emphasis was on investigating
the
spectral, modulation and subharmonic stability properties of the
dnoidal,
cnoidal and snoidal waveforms present in the model.
As an interesting side-product of the relevant investigation, we
explored
the dynamical evolution of unstable waveforms and observed
that they can lead to spatio-temporally localized structures in our
direct numerical simulations which are certainly worthwhile of further
investigation. Using the Hill's method, we have concluded the following modulational stability and instability results.
In the superluminal regime, the dnoidal waves are modulationally unstable and the spectrum forms a figure eight intersecting at the origin, while the cnoidal waves are modulationally stable and the spectrum near the origin is represented by the vertical bands along the purely imaginary axis. In the subluminal regime, the snoidal waves are modulationally unstable since the spectrum forms a figure eight intersecting at the origin. 
Considering the subharmonic perturbations, we have shown that the snoidal waves in the subluminal regime are spectrally unstable with respect to
all subharmonic perturbations including co-periodic ones (in line with
earlier work), while for dnoidal and cnoidal waves in the superluminal
regime, the transition between the spectrally stable state and the
spectrally unstable state occurs.  For cnoidal and dnoidal waves,
instability arises when two imaginary eigenvalues collide along the
imaginary axis in a Hamiltonian Hopf bifurcation and enter the right
and left half planes along the figure eight. We have also explained why
such
a bifurcation feature can only be present for traveling waves and is
demonstrably absent for stationary waves within this model.
For snoidal waves, on the other hand, we showed
that there are two eigenvalues (already for $P=1$) that were fixed on the real axis, which implies that the snoidal waves in the subluminal regime are spectrally unstable with respect to subharmonic perturbations. 


We note that our instability results are in accordance with the
  theory of nonlinear Klein-Gordon equation developed in~\cite{maramiller}, which
  tackles
  the Hamiltonian-Hopf instabilities of periodic traveling waves in
  such models. Our more concrete and detailed findings are tailored to
  the $\phi^4$
  case, yet they are in line with the more general theoretical
  findings of the above work, which considers the possibility of
  Hopf bifurcations
  for arbitrary Klein-Gordon equations and performs a local analysis
  of
  the spectrum near the imaginary axis.
Naturally, this study paves the way for numerous additional
computational developments. One important element
of our findings is the spontaneous emergence of spatio-temporally
localized
waves which we termed  localized waves on spatio-temporal backgrounds (LWSTs).
One could
seek such waves as exact numerical solutions using, e.g., methods such
as those of~\cite{cory}. Another possibility could be to seek to
generalize
such findings to higher-dimensional settings, which have also been
of substantial interest recently; see, e.g.,~\cite{ricardo} for a
relevant example. Such studies are presently in progress and will be
reported in future publications.

\end{document}